\documentclass[sigconf]{acmart} 
\AtBeginDocument{%
  }

\newcommand{\blue}[1]{{\color{blue} #1}}
\newcommand{\red}[1]{{\color{red} #1}}
\newcommand{\myparagraph}[1]{{\noindent \bf #1}}
\usepackage{color}
\usepackage{amsmath}
\usepackage{amsfonts}
\usepackage{amsthm}
\usepackage{mathtools}
\usepackage{bbm}
\usepackage{algorithm}
\usepackage{algpseudocode}
\usepackage{booktabs}      
\usepackage{pgfplotstable} 
\usepackage{pdflscape}     
\usepackage{siunitx}       
\pgfplotsset{compat=1.18}
\algrenewcommand\algorithmicrequire{\textbf{Input:}}
\algrenewcommand\algorithmicensure{\textbf{Output:}}

\newcommand{\secpar}{\ensuremath{\lambda}}

\newcommand{\ppt}{\mathsf{PPT}}
\newcommand{\commit}{\mathsf{Commit}}

\newcommand{\setup}{\mathsf{Setup}}
\newcommand{\open}{\mathsf{Open}}
\newcommand{\NN}{\mathbb{N}}

\newcommand{\Func}{\mathcal{F}}
\newcommand{\abort}{\mathsf{abort}}
\newcommand{\hon}{\mathcal{H}}
\newcommand{\mal}{\mathcal{M}}

\newcommand{\function}{\mathcal{F}}

\newcommand{\prot}{\Pi}

\newcommand{\prover}{\mathcal{P}}
\newcommand{\verifier}{\mathcal{V}}

\newcommand{\poly}{\mathsf{poly}}

\newcommand{\adversary}{\mathcal{A}}

\newcommand{\hybrid}{\mathcal{H}}

\newcommand{\bits}{\{0,1\}}

\newcommand{\ignore}[1]{}

\newcommand{\com}{\ensuremath{\mathtt{com}}}

\newcommand{\simu}{\mathsf{Sim}}

\newcommand{\NP}{NP}

\newcommand{\negligible}{\nu}

\newcommand{\remove}[1]{}

\newcounter{itemcount4}

\newcommand{\Sim}{\mathcal{S}}

\newcommand{\ideal}{\mathsf{IDEAL}}

\newcommand{\real}{\mathsf{REAL}}

\newcommand{\crs}{\mathtt{CRS}}

\newcommand{\setupcom}{\mathsf{SetupCom}}
\newcommand{\setupzkp}{\mathsf{SetupZKP}}

\newcommand{\relation}[1]{\mathcal{R}_{#1}}

\newcommand{\sample}{\overset{\$}{\gets}}

\newcommand{\alice}{P_1}
\newcommand{\bob}{P_2}

\newcommand{\ppnew}{\mathsf{pp}}
\newcommand{\arrowvec}[1]{\overrightarrow{#1}}

\newcommand{\score}{\mathsf{Score}}

\newcommand{\mtsetup}{\mathsf{mtSetup}}
\newcommand{\mtcommit}{\mathsf{mtCommit}}
\newcommand{\mtopen}{\mathsf{mtOpen}}
\newcommand{\mtverify}{\mathsf{mtVerify}}

\usepackage{booktabs}
\usepackage{enumitem}
\usepackage{latexsym}
\usepackage{mathtools}
\usepackage{multirow}
\usepackage{multicol}
\usepackage{mwe}
\usepackage{makecell}
\usepackage{pgfplots}
\usepackage{pifont}
\usepackage{ragged2e}
\usepackage{rotating}
\usepackage{subcaption}
\usepackage{tabu}
\usepackage{tabularx}
\usepackage{threeparttable}
\usepackage{tikz}
\usepackage[normalem]{ulem}
\usepackage{url}
\usepackage{verbatim}
\usepackage{wrapfig}
\usepackage{xcolor}
\usepackage{xspace}
\usepackage[framemethod=tikz]{mdframed}
\usepackage{algpseudocode}
\usepackage[most]{tcolorbox}

\usepackage{amsfonts,amsmath,amssymb,amsthm}
\usepackage{multirow}
\usepackage{hyperref}
\usepackage{svg}
\usepackage{float}
\usepackage[
    n,
    operators,
    advantage,
    sets,
    adversary,
    landau,
    probability,
    notions,    
    logic,
    ff,
    mm,
    primitives,
    events,
    complexity,
    asymptotics,
    keys]{cryptocode}
\usepackage{algorithm}
\algrenewcommand\algorithmicrequire{\textbf{Input:}}
\algrenewcommand\algorithmicensure{\textbf{Output:}}

\usepackage{breakcites}

\expandafter\def\csname ver@etex.sty\endcsname{}

\makeatletter
\def\th@plain{%
  \thm@notefont{}
  \itshape 
}
\def\th@definition{%
  \thm@notefont{}
  \normalfont 
}
\makeatother

\usepackage{mdframed}
\newmdenv[%
  backgroundcolor=gray!10,
  linecolor=black,
  linewidth=1pt,
  nobreak=true,
  innerleftmargin=5pt,
  innerrightmargin=10pt,
  innertopmargin=1pt,
  innerbottommargin=10pt,
  font=\normalfont,
]{theorembox}
\input{symbols_env}

\usepackage{booktabs}
\usepackage[table]{xcolor}
\usepackage{tabularx}
\definecolor{SplitA}{HTML}{E3F2FD} 
\definecolor{SplitB}{HTML}{FFF3E0} 
\definecolor{SplitC}{HTML}{E8F5E9} 
\newcommand{\bandA}{\cellcolor{SplitA}}
\newcommand{\bandB}{\cellcolor{SplitB}}
\newcommand{\bandC}{\cellcolor{SplitC}}

\copyrightyear{2026}
\acmYear{2026}
\setcopyright{cc}
\setcctype{by}
\acmConference[ASIA CCS '26]{ACM Asia Conference on Computer and Communications Security}{June 01--05, 2026}{Bangalore, India}
\acmBooktitle{ACM Asia Conference on Computer and Communications Security (ASIA CCS '26), June 01--05, 2026, Bangalore, India}
\acmDOI{10.1145/3779208.3785393}
\acmISBN{979-8-4007-2356-8/2026/06}

\begin{document}

\title{PrivaDE: Privacy-preserving Data Evaluation for Blockchain-based Data Marketplaces}

\author{Wan Ki Wong}
\email{thomas.wong@ed.ac.uk}
\affiliation{%
  \institution{University of Edinburgh}
  \city{Edinburgh}
  \country{United Kingdom}
}

\author{Sahel Torkamani}
\email{sahel.torkamani@ed.ac.uk}
\affiliation{%
  \institution{University of Edinburgh}
  \city{Edinburgh}
  \country{United Kingdom}
}

\author{Michele Ciampi}
\email{michele.ciampi@ed.ac.uk}
\affiliation{%
  \institution{University of Edinburgh}
  \city{Edinburgh}
  \country{United Kingdom}
}

\author{Rik Sarkar}
\email{rik.sarkar@ed.ac.uk}
\affiliation{%
  \institution{University of Edinburgh}
  \city{Edinburgh}
  \country{United Kingdom}
}

\renewcommand{\shortauthors}{Wong et al.}

\begin{abstract}
Evaluating the usefulness of data before purchase is essential when obtaining data for high-quality machine learning models, yet both model builders and data providers are often unwilling to reveal their proprietary assets. 

We present PrivaDE, a privacy-preserving protocol that allows a model owner and a data owner to jointly compute a utility score for a candidate dataset without fully exposing model parameters, raw features, or labels. PrivaDE provides strong security against malicious behavior and can be integrated into blockchain-based marketplaces, where smart contracts enforce fair execution and payment. To make the protocol practical, we propose optimizations to enable efficient secure model inference, and a model-agnostic scoring method that uses only a small, representative subset of the data while still reflecting its impact on downstream training. Evaluation shows that PrivaDE performs data evaluation effectively, achieving online runtimes within 15 minutes even for models with millions of parameters.

Our work lays the foundation for fair and automated data marketplaces in decentralized machine learning ecosystems.
\end{abstract}



\maketitle

\section{Introduction}
\label{sec:introduction}

As machine learning (ML) systems proliferate and generate substantial value, concerns have grown over large-scale, uncompensated data collection and the need to return value to original contributors. Collaborative learning frameworks~\cite{collabML}, in which multiple parties pool data via a decentralized ledger, offer one remedy: trainers publish data requests, contributors opt in, and contributions are compensated in a transparent and fair manner. Despite the rise of blockchain-based data and ML marketplaces (e.g., \cite{rao_bittensor_whitepaper}), rigorous mechanisms for data evaluation remain largely overlooked.

The central challenge of data evaluation in collaborative learning is how to select high-quality samples that most improve model performance while preserving the confidentiality of both data and model assets. The problem is magnified in decentralized, privacy-sensitive marketplaces, where neither contributors nor model owners are willing to reveal raw features, labels, or model parameters prior to payment. It is therefore essential to design cryptographically secure protocols that compute data-utility scores under strong confidentiality guarantees while ensuring correctness in a trustless environment.

In this paper, we design a secure protocol that enables a model owner (MO) and a data owner (DO) to jointly compute a data-utility score $\phi(M,D)$ for a model $M$ and dataset $D$, with practical applicability to blockchain-based data marketplaces. This score can subsequently serve as the basis for negotiation or automated payment for data acquisition in such marketplaces.

\myparagraph{Our Contributions.}
We introduce \textsc{PrivaDE}, a two-party, maliciously secure protocol for dataset scoring. \textsc{PrivaDE} has two main components: (i) an improved and more efficient secure model inference protocol that obtains the model's predictions on the dataset, and (ii) a data scoring protocol that takes the data points and model predictions as input and outputs a score $\phi$ representing the usefulness of the data.

Secure model inference under malicious security, e.g., with SPDZ \cite{mpspdz} or MASCOT \cite{mascot}, is known to incur very high overhead. We introduce three optimization techniques (Section~\ref{sec:design}) that make the secure inference component of \textsc{PrivaDE} practical under a strict malicious-adversary assumption. In particular, we combine model distillation (to shrink the model), model splitting (to reduce the portion of computation run in secure environments), and a cut-and-choose ZKP audit (to reduce verification cost) to carefully balance efficiency and security.

We further design a novel, model-agnostic scoring function (Alg.~\ref{alg:multi-public}) for the data scoring component, which evaluates data points and yields a score that correlates with downstream training performance. The scoring function builds on active learning criteria, namely diversity (how spread out the points are) and uncertainty (how uncertain the model's predictions are), which have been empirically shown to be effective heuristics for data quality. Our scoring function is easy to compute under maliciously secure protocols and minimizes passive information leakage by operating on a \emph{representative subset} of the dataset. The scoring procedure is constructed so that the subset-derived score accurately reflects the full dataset.

We provide a brief proof-of-concept integration of \textsc{PrivaDE} into a blockchain-based data marketplace. Smart contracts deter misuse and enforce fairness in MPC by penalizing uncooperative parties via escrow slashing, and the protocol can be tightly coupled with the data transaction to streamline the end-to-end marketplace workflow.

Finally, we prove security in the simulation-based paradigm and demonstrate practicality through extensive experiments (Section~\ref{sec:experiments}) on real-world datasets and standard model architectures. Our results show that \textsc{PrivaDE} can be flexibly tuned to achieve reasonable runtimes (as low as 45 seconds of online time for MNIST with LeNet), and that the scoring protocol is robust, effectively identifying high-value datasets.

\begin{figure}[htbp]
    \centering
    \includegraphics[width=\columnwidth]{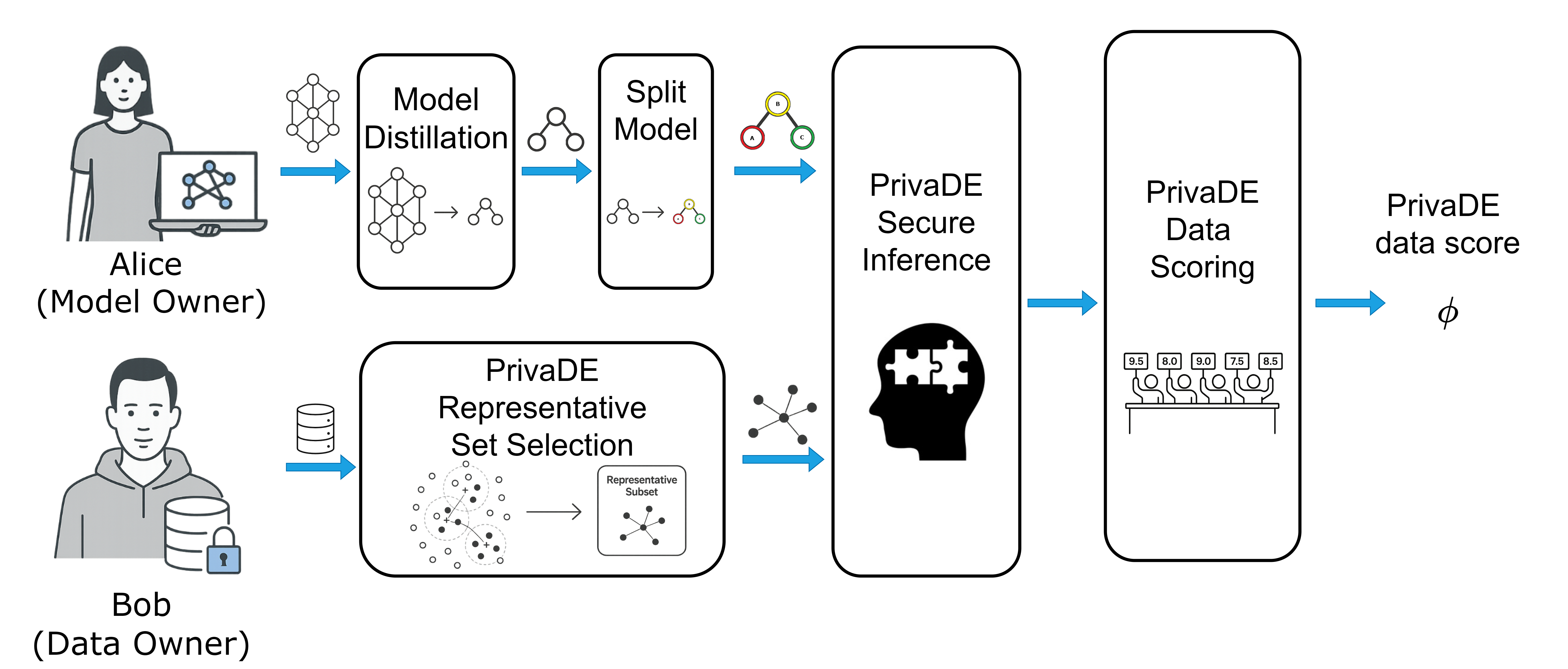}
    \caption{Overall design of \textsc{PrivaDE}. The core protocol consists of two parts: a secure inference of the data points and a scoring function that takes the inference results and labeled data points as input. Several optimization components, detailed in Section~\ref{sec:design}, are applied before secure inference to reduce the overhead of the first component.}
    \Description{A flowchart showing the overall design and workflow of PrivaDE.}
    \label{fig:privade}
\end{figure}

To our knowledge, this is the first work to propose a secure data evaluation scheme in the malicious setting together with a model-agnostic, crypto-friendly scoring function. Prior work (e.g., \cite{songDE,qianDE,privatedv}) either operates in the semi-honest model or relies on a trusted third party (TTP) for most protocol steps. An overview of \textsc{PrivaDE} is shown in Fig.~\ref{fig:privade}.

\paragraph{Organization.}
Section~\ref{sec:preliminaries} reviews background on machine
learning, cryptography, and blockchain. Section~\ref{sec:problem}
formalizes the problem and threat model.  Section~\ref{sec:protocol} presents the full protocol and security proofs. 
Section~\ref{sec:design} explains optimization techniques applied that reduce the overhead of secure inference.
Section~\ref{sec:blockchain} describes a proof-of-concept integration of \textsc{PrivaDE} into a blockchain-based marketplace, Section~\ref{sec:experiments} evaluates the
practicality and robustness of our construction, and
Section~\ref{sec:related} discusses related work.

\section{Preliminaries}
\label{sec:preliminaries}

\subsection{Machine learning}
In supervised machine learning, we compute a model $h$ from a training set $D=\{(x_i,y_i)\}_{i=1}^n$, where each $x_i$ has label $y_i$ (sometimes $z_i=(x_i,y_i)$). An algorithm $A$ searches a hypothesis space $\mathcal{H}$ for $h$ minimizing empirical loss $L_D(h)=\frac{1}{n}\sum_{i=1}^n \ell\bigl(h(x_i),y_i\bigr)$,
where $\ell$ quantifies the error between prediction $y'=h(x)$ and true $y$. Common choices include squared loss $\ell(y',y)=(y'-y)^2$ for regression and zero-one loss $\ell(y',y)=\mathbf{1}[y'\neq y]$ for classification. For probabilistic outputs $\vec y'$, cross-entropy $\ell_{\mathrm{CE}}(\vec y',\vec y)=-\sum_{i=1}^d \vec y^i\ln\vec y'^i$ is typical, where $\vec y$ is a one-hot vector. We denote models by $M$ and, when parameterized by $\theta$, by $M_\theta$.

\subsection{Cryptographic protocols}
This section reviews commitment schemes, zero-knowledge proofs (ZKPs), and multi-party computation (MPC). We use $\lambda$ as the security parameter and assume familiarity with computational indistinguishability and negligible functions $\negligible$.

\subsubsection{Commitment schemes}
\label{def: commitment}
Informally, a commitment scheme is a tuple $\Gamma = (\setup,\allowbreak \commit, \open)$ of $\ppt$ algorithms where: 
\begin{itemize} [itemsep=0em,nosep]
        \item $\setup(1^{\lambda}) \rightarrow \ppnew$ takes security parameter $\lambda$ and generates public parameters $\text{pp}$;
        \item $\commit(\ppnew; m) \rightarrow (\com,r)$ takes a secret message $m$, outputs a public commitment $Com$ and a randomness $r$ used in the commitment.
        \item $\open(\ppnew, \com; m, r) \rightarrow b \in \{0,1\}$ verifies the opening of the commitment $C$ to the message $m$ with the randomness $r$.
\end{itemize}

We require two main security properties: \emph{hiding} and \emph{binding}. Hiding ensures no $\ppt$ adversary can determine the committed message from $\com$. Binding ensures $\com$ can be opened to only one specific message. See App.~\ref{app:com} for formal definitions.

\subsubsection{Merkle Tree Commitments}
A Merkle tree commitment uses a collision-resistant hash function $H$ to commit to a vector $(m_1,$ $\ldots,$ $m_n)$ by hashing leaves and iteratively hashing pairs up a binary tree; the commitment is the root, and an opening is the $O(\log n)$-length authentication path for an index $i$.

A merkle tree commitment is a tuple of PPT algorithms $\Gamma = (\mtsetup,\mtcommit,\mtopen,\mtverify)$ where:
\begin{itemize}[leftmargin=1.25em]
  \item $\mtsetup(1^\lambda)\to\ppnew$: generates public parameters $\ppnew$./
  \item $\mtcommit(\ppnew; m_1,\ldots,m_n)\to(\com,\st)$: compute leaf and internal hashes; output root $\com$ and state $\st$.
  \item $\mtopen(\ppnew,\st; i)\to\pi_i$: output the sibling-hash path authenticating position $i$.
  \item $\mtverify(\ppnew,\com; i,m_i,\pi_i)\to b$: recompute the root from $(i,$ $m_i,$ $\pi_i)$ and accept iff it equals $\com$.
\end{itemize}

The position-binding property (no two different $m$ open at the same $i$) holds under collision resistance of $H$. We refer to App.~\ref{app:merkle} for the formal definition.

\subsubsection{Zero-knowledge proofs}
\label{def: NIZK}
A {\em non-interactive zero-knowledge proof-of-knowledge system (NIZKPoK)} for an \NP-language $L$ with the corresponding relation $\relation{L}$ is represented by a tuple of algorithms $\Pi = (\setup, \prover, \verifier)$, where:
    \begin{itemize}[itemsep=0em,nosep]
        \item $\setup( 1^{\lambda}) \rightarrow \crs$ takes as the input a security parameter $\lambda$. It outputs a common reference string $\crs$.
        \item $\prover(\crs, x, w) \rightarrow \pi$ takes as the input $\crs$, the statement $x$ and the witness $w$, s.t. $(x, w) \in \relation{L}$. It outputs the proof $\pi$.
        \item $\verifier(\crs, x, \pi) \rightarrow b \in \{0,1\}$ takes as the input $\crs$, $x$ and $\pi$. It outputs 1 to accept and 0 to reject.
    \end{itemize}

Informally, a NIZK proof is secure if it satisfies \emph{completeness}, \emph{zero-knowledge}, and \emph{soundness}. Completeness requires honest verifiers always accept honest proofs. Zero-knowledge ensures proofs leak no information about the witness $w$, even to malicious verifiers. Soundness requires verifiers reject proofs of false statements.
We refer to App.~\ref{app:zk} for the formal definition.

\subsubsection{Multi-party computation}
\label{def: mpc}
We prove that our results are secure under the simulation-based paradigm of secure \emph{Multi-Party Computation (MPC)}. In secure multiparty computation (MPC) multiple parties aim to compute a function $f$ together while keeping their inputs private. The goal is to ensure the result is correct while revealing no extra information.
We say that a protocol is secure if whatever can be inferred by an adversary attacking the protocol can also be inferred by an adversary attacking a so-called \emph{ideal-world}. In this, there is an ideal trusted third party that receives the inputs of the parties evaluates the function $f$, and returns the output of the computation back to the parties 

In short, a protocol is secure if a real-world adversary cannot gain an advantage over what they would learn in the ideal-world setting (that trivially captures the best possible security guarantee). 
When defining the security of a protocol $\Pi$, we will say that $\Pi$ \emph{securely realizes} the function $f$, if $\Pi$ guarantees the above.
In this paper we focus on the two-party setting, considering security against \emph{malicious} adversaries. Parties can deviate arbitrarily from what the protocol prescribes. 
In App.~\ref{app:mpc} we provide a formal definition of secure MPC and its variants.

\subsection{Blockchain and Smart Contracts}

A \emph{blockchain} is a decentralized, append-only ledger storing an ordered sequence of blocks; each block includes transactions, a timestamp, and the previous block's hash, yielding immutability~\cite{nakamoto2008bitcoin,bonneau2015sok}. A \emph{smart contract} is a program replicated and executed by blockchain nodes, enforcing state transitions without intermediaries~\cite{buterin2014next,wood2014ethereum}.

PrivaDE targets an EVM-compatible chain, which provides a secure, verifiable substrate for data exchange and computation. We use the chain to:
\begin{itemize}
  \item \textbf{Commit inputs.} Data contributions and model parameters are posted as cryptographic commitments, creating a binding, immutable audit trail.
  \item \textbf{Enforce MPC fairness.} Contracts escrow deposits and impose penalties for deviations, discouraging aborts and enforcing compliance during MPC \cite{fairmpc}.
  \item \textbf{Enable atomic payment.} Transfers to data contributors trigger only on verifiable completion (e.g., successful checks/proofs), preventing unilateral defection.
\end{itemize}

\section{Problem Description and Threat Model}
\label{sec:problem}
Consider a scenario in which {\em Alice} owns a model $M$ trained on her dataset $D_A$ and seeks additional data to further improve it. {\em Bob} holds a relevant dataset $D_B$ that could enhance $M$ when combined with $D_A$ (i.e., $D_A \cup D_B$). Bob seeks fair compensation for his contribution, while Alice aims to pay in proportion to the improvement attributable to $D_B$. We assume $D_B$ is authentic and pertinent to Alice’s task; authenticity can be attested, for example, via signatures by the original data source or a trusted authority (Sec.~\ref{sec:blockchain}).

Our objective is to compute a score $\phi=\score(M,D_B)$ that quantifies the improvement $D_B$ affords to $M$, approximating the net gain in model performance (e.g., increased accuracy on a held-out test set). In pursuit of this goal, we address two coupled questions:
\begin{enumerate}
\item What criterion should define the scoring function $\score$?
\item How can we compute $\score$ securely while protecting both Alice's and Bob's private data?
\end{enumerate}
The challenge is to devise a criterion that reliably reflects true improvement—ordinarily requiring retraining—while remaining simple enough for efficient secure computation within practical runtimes. Consequently, we co-design the scoring rule and the secure protocol to yield a feasible and efficient scoring system.

We assume Bob contributes a dataset with multiple points ($|D_B|>1$), reflecting practical scenarios such as a company representative contributing bulk records (e.g., sales data). To facilitate deployment on blockchain-based systems, PrivaDE is maliciously secure: it executes in the presence of a PPT adversary $\adversary$ who may arbitrarily deviate from the protocol, with the identity of the corrupted party fixed prior to execution. The protocol protects the honest party by detecting any deviation and safely aborting. PrivaDE requires no trusted party beyond the setup and preprocessing stages.

PrivaDE assumes the model architecture is public, while the model weights $\theta$ are Alice's private input. Bob's private input is the dataset $D_B$, formally represented by the sequences $(\arrowvec{x},\arrowvec{y})$. The $i$-th sample is $(x_i,y_i)$, the $i$-th entries of $\arrowvec{x}$ and $\arrowvec{y}$.

The primary goal of our protocols is to protect $\theta$ and $(\arrowvec{x},\arrowvec{y})$. For efficiency, we allow carefully bounded, protocol-specified leakage, supported by prior work and experiments, which does not reveal the full model $\theta$ or any individual sample $(x_i,y_i)$. The remaining, unleaked components are proven secure via formal, simulation-based arguments; see Sec.~\ref{sec:protocol} for definitions and proofs.

\section{PrivaDE: The Data Evaluation Protocol}
\label{sec:protocol}
In this section, we describe the end-to-end workflow of \textsc{PrivaDE}.

\subsection{Dataset Scoring Algorithm}
As proven in numerous active-learning literature (e.g. \cite{settles2009active,dasgupta2011two}), the use of active learning heuristics often gives good indicator of the overall dataset quality. Hence, given a representative set $D_R$, the parties compute a score using three criteria - diversity, uncertainty, and loss, where diversity and uncertainty follow standard active-learning practice.

\myparagraph{Diversity.}
The subset should be diverse, with elements sufficiently different to avoid redundancy. We denote diversity by $\mathcal{D}$, instantiated as a statistic (mean, median) of pairwise distances, common in coreset and batch active learning~\cite{sener2018activelearningconvolutionalneural}.

\myparagraph{Uncertainty.}
Uncertainty indicates potential informativeness: high uncertainty suggests the model benefits from training on such samples. Typical measures include entropy of $M$'s predictive distribution or margin-based variants. We write the uncertainty functional as $\mathcal{U}$.

\myparagraph{Loss.}
Relevance is captured by loss $M$ incurs on Bob's data~\cite{yoo2019learning}. If $y'=M_\theta(x)$, then $\ell(y',y)$ quantifies prediction-label discrepancy; larger loss indicates greater potential training value.

Let
\[
\begin{aligned}
l &= \frac{1}{|D_R|} \;\; \sum_{\mathclap{(x,y)\in D_R}} \ell\!\left(M_\theta(x),y\right),\\
u &= \mathcal{U}\!\left(M_\theta[D_R]\right),\\
d &= \mathcal{D}\!\left(D_B^x, D_R^x\right).
\end{aligned}
\]
The overall score is $\phi = f(l,u,d)$ for a pre-agreed aggregator $f$. Algorithm~\ref{alg:multi-public} summarizes the cleartext procedure, where $D_B^x=\{x\mid(x,y)\in D_B\}$ and $D_B^y=\{y\mid(x,y)\in D_B\}$.

\begin{algorithm}
\caption{Cleartext dataset scoring $\score_\mathsf{multi}$}\label{alg:multi-public}
\begin{algorithmic}[1]
\Require model $M_\theta$, dataset $D_B$, loss $\ell$, uncertainty $\mathcal{U}$, diversity $\mathcal{D}$, scoring $f$, representative sampler $\mathcal{R}$
\Ensure score $\phi$ of $D_B$
\State $D_R \gets \mathcal{R}(D_B,k)$
\State For each $(x_i,y_i)\in D_R$, compute $y'_i \gets M_\theta(x_i)$
\State $l \gets \tfrac{1}{k}\sum_i \ell(y'_i,y_i)$;\quad $u \gets \mathcal{U}(\{y'_i\}_{i=1}^k)$;\quad $d \gets \mathcal{D}(D_B^x,D_R^x)$
\State $\phi \gets f(l,u,d)$
\State \Return $\phi$
\end{algorithmic}
\end{algorithm}

\paragraph{Example.}
For multi-class image classification, one can choose cross-entropy loss $\ell_{\mathrm{CE}}$, margin-based uncertainty $u=\frac{1}{k}\sum_i(\max p_i-\text{second-max }p_i)$ \cite{settles2009active}, diversity $d=$ mean pairwise $\|x-x'\|$ over $D_R^x$, and $f(l,u,d)=\alpha l+\beta u+\gamma d$ for weights $\alpha,\beta,\gamma$.

The exact choices of $\ell, \mathcal{U}, \mathcal{D}, f$ are decided case-by-case. For example, an untrained model may benefit from higher diversity, while a fine-tuned model may benefit from higher uncertainty or loss. The model owner tunes the scoring function to their needs; our protocol construction and security are oblivious to this choice. 

\subsection{Supplementary Components}
\label{subsec:building-blocks}
Before presenting the full protocol, we briefly introduce three supplementary components of \textsc{PrivaDE}, shown in Fig.~\ref{fig:privade} preceding the two core components. These components form part of the optimization techniques of \textsc{PrivaDE}, which are described in greater detail in Sec.~\ref{sec:design}.

\myparagraph{Model Distillation.}
The model owner may optionally perform distillation on their model $M$, training a smaller model that closely approximates the performance of the original. In the remainder of this section, we assume Alice's model $M$ denotes this distilled (smaller) model. See Sec.~\ref{design:md} for more details.

\myparagraph{Split Model.}
We require Alice, the model owner, to partition the model \(M\) into three blocks: \(M = C \circ B \circ A\), where block \(B\) is revealed to Bob while \(A\) and \(C\) remain private to Alice. This design reduces cryptographic overhead while balancing the leakage of the protocol. The detailed security analysis and splitting methodology are presented in Sec.~\ref{design:sm}.

\myparagraph{Representative Set Selection.}
We require Bob to select a \emph{representative subset} $D_R \subset D_B$ before running the core protocol, specifically a $(d,\delta)$\emph{-representative} subset:
\begin{definition}[Representativeness of $D_R$ to $D_B$]
\label{def:rep}
Let $D_R\subset D_B$, and fix a distance threshold $d>0$ and outlier tolerance $\delta\in[0,1]$. Then $D_R$ is $(d,\delta)$\emph{-representative} if
\[
\Pr_{x\sim D_B}\!\bigl[\exists\,x'\in D_R:\|x-x'\|<d\bigr]\ge 1-\delta,
\]
where the probability is taken with respect to the empirical distribution over $D_B$.
\end{definition}

We write $\mathcal{R}(D_B)$ for any algorithm that, given $|D_B|=n>k$, returns $D_R\subset D_B$ of size $k$. Bob may use any appropriate algorithm $\mathcal{R}$; we only verify the representativeness of the selected subset.

To verify representativeness, we propose a \emph{challenge protocol} $\mathsf{CP}$. Bob commits each $x_i\in D_B^x$ via $\com_1,\dots,\com_n$ and publishes the representative set $D_R$ by revealing indices $I_R := \{i | x_i \in D_R^x\}$.
Alice samples random $i\in[n]$ and challenges Bob to prove in zero knowledge that committed $x_i$ is well represented by $D_R$. Bob recovers $x_i$, computes $d_i'=\min_{x^r_j\in D_R}\|x^r_j-x_i\|$, and:
\begin{itemize}
  \item if $d_i'<d$, returns a ZK proof $\pi_i$ attesting that
  $\exists j,r_i:\ \|x^r_j-x_i\|\le d\ \wedge\ \open(\ppnew,\com_i,x_i,r_i)=1$;
  \item otherwise, replies $\mathsf{fail}_i$.
\end{itemize}
By repeating this challenge, Alice gains confidence that $D_R$ is representative of $D_B$. A detailed analysis of representative set selection and the challenge protocol is given in Sec.~\ref{design:rss}.

\subsection{The \textsc{PrivaDE} Protocol}
\label{subsec:privade}

We now give a secure realization of Algorithm~\ref{alg:multi-public}. At a high level, the protocol proceeds in four stages:
\begin{enumerate}[label=(\roman*)]
    \item \textbf{Setup.} Public parameters, keys, and any preprocessing required by the underlying secure primitives are initialized.
    \item \textbf{Representative selection \& audit.} Bob computes a representative set $D_R$, and Alice audits it using the challenge protocol $\mathsf{CP}$ (Sec.~\ref{subsec:CP}) to ensure $(d,\delta)$-representativeness on sufficiently many points.
    \item \textbf{Verified inference.} The model is evaluated on $D_R$, and a corresponding zero-knowledge proof is used to verify computation. 
    \item \textbf{Secure scoring.} Alice and Bob invoke a maliciously secure 2PC that realizes $\Func_\mathsf{SubScore}$ (Fig.~\ref{fig:func-dec-multi}) to compute the final score from loss, uncertainty, and diversity statistics.
\end{enumerate}

Formally, we prove that the protocol realizes $\Func^k_{\score}$, which securely computes Algorithm~\ref{alg:multi-public}. In particular, Alice’s input (the model weights) and Bob’s input (dataset) are protected except for some intended leakage, to which the degree is primarily tuned by the parameter $k$ (size of the representative set). This leakage profile is fully specified by $\Func^k_{\score}$ in Fig.~\ref{fig:func-val2}.

\begin{functionalitysplitbox}{$\Func^k_{\score}$}{Functionality \(\Func^k_{\score}\) for computing the score of a dataset.}{fig:func-val2}
\begin{itemize}
    \item[] \textbf{Parameters:} model \(M = C \circ B \circ A\), loss \(\ell\), uncertainty function \(\mathcal{U}\), diversity function \(\mathcal{D}\), scoring function \(f\), representative set selection function \(\mathcal{R}\), represenattive set size $k$, distance threshold \(d\), outlier ratio \(\delta\), commitment parameter $\ppnew$
    \item[] 
    \begin{enumerate}[itemsep=0em,nosep]
    \item $\Func^k_{\score}$ receives $I_R$ of size $k$ (indices of representative set) from $\bob$.
        \item $\Func^k_{\score}$ receives a list of challenge indices $I$ from $\alice$, and sends $I$ to $\bob$.
        \item $\Func^k_{\score}$ receives $D^x_{B|I}$ (The set $D_B^x$ indiced by $I$) and $D^x_{R|I}$ (respective points in $D^x_R$ closest to each $D^x_{B|I}$) from $\bob$. For each \(i \in I\) $(x_i \in D^x_{B|I})$, $\Func^k_{\score}$ checks that there exists some \(x_j \in D_{R|I}^x\) such that \(\|x_i - x_j\| < d\). If this check fails for at least \(\delta \cdot |I|\) indices, the functionality sends \(\abort\) to both \(\alice\) and \(\bob\).
        \item $\Func^k_{\score}$ receives $\theta_A$ from $\alice$ and $D^x_R$ (of size $k$) from $\bob$, then computes $\arrowvec{a_A} \gets A_{\theta_A}(D^x_R)$ and sends $\arrowvec{a_A}$ to  $\bob$.
        \item $\Func^k_{\score}$ receives $\theta_B$ from $\alice$, sends $\theta_B$ to $\bob$, then computes $\arrowvec{a_B} \gets B_{\theta_B}(\arrowvec{a_A})$ and sends $\arrowvec{a_B}$ to  $\alice$ and $\bob$.
        \item $\Func^k_{\score}$ receives $\theta_C$ from $\alice$, computes $\arrowvec{y'} \gets C_{\theta_C}(\arrowvec{a_B})$.
        \item $\Func^k_{\score}$ receives $D^x_{B}, D^y_R$ from $\bob$ and calculates the score: $l = \frac{1}{k}\sum_{i=1}^{k}\ell(y'_i,y_i), \quad u = \mathcal{U}(\{y'_1,\ldots,y'_k\}), \quad d = \mathcal{D}(D_B^x,D_R^x)$,
        and finally \(\phi \leftarrow f(l,u,d)\). It then sends \(\phi\) to both \(\alice\) and \(\bob\).
    \end{enumerate}
\end{itemize}
\end{functionalitysplitbox}

We are now ready to formally describe the protocol \textsc{PrivaDE} that realizes \(\Func^k_{\score}\) (Fig.~\ref{fig:func-val2}). \textsc{PrivaDE} uses the following tools:

\begin{itemize}[itemsep=0em,nosep]
    \item \((\setupcom,\commit,\open)\) is a commitment scheme satisfying Definition~\ref{def: commitment}.
    \item \((\setupzkp,\prover,\verifier)\) is an NIZKPoK system satisfying Definition~\ref{def: NIZK}, defined for the NP language \(L_\textsf{ML-Inf}\). We consider two scenarios: \blue{(1) the model weights and outputs are hidden as part of the witness and the data point is public (blue)}; \red{(2) the data point is hidden as part of the witness and the model weights and outputs are public (red)}. For each scenario, read the black text together with the text in the corresponding colour.
    \[
L_{\textsf{ML-Inf}} = \left\{\,
  \begin{array}{@{}l@{}}
    (\ppnew,
     \blue{\arrowvec{x},\com_\theta,\arrowvec{\com_{y'}}},
     \red{\theta,\arrowvec{\com_x},\arrowvec{y'}}):\\[0.2em]
    \exists\,(\blue{\theta,r_\theta,\arrowvec{y'},\arrowvec{r_{y'}}},
              \red{\arrowvec{x},\arrowvec{r_x}}) \ \text{s.t.} \\
    M_\theta(x)=y', \\
    \blue{\open(\ppnew,\com_\theta,\theta,r_\theta)=1}, \\
    \blue{\forall y' \, \open(\ppnew,\com_{y'},y',r_{y'})=1}, \\
    \red{\forall x \,\open(\ppnew,\com_x,x,r_x)=1}
  \end{array}
\right\}.
\]

    Our proof-of-concept implementation uses EZKL~\cite{zkonduit_ezkl}.
    \item \(\Pi_\mathsf{Inference}\) is a 2-party protocol satisfying Definition~\ref{def: mpc}, which securely realizes \(\Func_\mathsf{Inference}\) (Fig.~\ref{fig:func-inf}) with malicious security. We instantiate this with SPDZ2k~\cite{Cramer2018SpdZ2k} in MP-SPDZ~\cite{mpspdz}.
    \item \(\Pi_\mathsf{SubScore}\) is a 2-party protocol satisfying Definition~\ref{def: mpc}, which securely realizes \(\Func_\mathsf{SubScore}\) (Fig.~\ref{fig:func-dec-multi}); also instantiated with SPDZ2k.
\end{itemize}
\begin{functionalitysplitbox}{$\Func_\mathsf{Inference}$}{Functionality \(\Func_\mathsf{Inference}\) for maliciously secure 2PC.}{fig:func-inf}
    \begin{itemize}
        \item[] \textbf{Parameters.} Description of the model $A$; model owner $\alice$; data owner $\bob$; commitments $\com_A, \arrowvec{\com_{x}}$; commitment parameter $\ppnew$.
    
        \item[] 
    \begin{enumerate}[itemsep=0em,nosep]
        \item \(\Func_\mathsf{Inference}\) receives private input $(\theta_A,r_A)$ from \(\alice\) and \((\arrowvec{x},\arrowvec{r_{x}})\) from \(\bob\).
        \item it verifies the commitments $
        b_{A} \leftarrow \open(\ppnew,\com_{A},\theta_A,r_{A}) \quad \text{and} \quad b_{x_i} \leftarrow \open(\ppnew,\com_{x_i},x_i,r_{x_i})$
        for all \(i\). If \(b_{A}\neq 1\) or \(b_{x_i} \neq 1\) for any \(i\), it sends \(\abort\) to both \(\alice\) and \(\bob\).
        \item It computes $\arrowvec{y'} \gets A(\arrowvec{x})$, commits $\com_{a_i}, r_{a_i} \gets \commit(\ppnew,y_i)$ for each $y_i \in \arrowvec{y'}$ 
        \item It delivers \(\arrowvec{y'}, \{\com_{a_i}, r_{a_i}\}_i\),  to \(\bob\), and only $\{\com_{a_i}\}_i$ to $\alice$.
    \end{enumerate}
    \end{itemize}
\end{functionalitysplitbox}

\begin{functionalitysplitbox}{$\Func_\mathsf{SubScore}$}{Functionality \(\Func_\mathsf{SubScore}\) for maliciously secure 2PC.}{fig:func-dec-multi}
    \begin{itemize}
        \item[] \textbf{Parameters.} Loss \(\ell\); uncertainty \(\mathcal{U}\); diversity \(\mathcal{D}\); aggregator \(f\); model owner \(\alice\); data owner \(\bob\); commitment parameter \(\ppnew\); public representative features \(D_R^x\).
    
        \item[] 
    \begin{enumerate}[itemsep=0em,nosep]
        \item \(\Func_\mathsf{SubScore}\) receives private input \((y'_1,\ldots,y'_k, r_{y'_1},\ldots,r_{y'_k}, \com_{y_1},\ldots,\com_{y_k})\) from \(\alice\) and \((D_B^x,y_1,\ldots,y_k, r_{y_1},\ldots,r_{y_k}, \com_{y'_1},\ldots,\com_{y'_k})\) from \(\bob\).
        \item It verifies the commitments $
        b_{y_i} \leftarrow \open(\ppnew,\com_{y_i},y_i,r_{y_i}) \quad \text{and} \quad b_{y'_i} \leftarrow \open(\ppnew,\com_{y'_i},y'_i,r_{y'_i})$
        for all \(i\). If \(b_{y_i}\neq 1\) or \(b_{y'_i} \neq 1\) for any \(i\), it sends \(\abort\) to both \(\alice\) and \(\bob\).
        \item It computes the average loss 
        \(
        l = \frac{1}{k}\sum_{i=1}^{k}\ell(y'_i,y_i),
        \)
        the uncertainty score 
        \(
        u = \mathcal{U}(\{y'_1,\ldots,y'_k\}),
        \) the diversity score 
        \(
        d = \mathcal{D}(D_B^x,D_R^x)
        \) and finally \(\phi \leftarrow f(l,u,d)\). It then outputs \(\phi\) to both \(\alice\) and \(\bob\).
    \end{enumerate}
    \end{itemize}
\end{functionalitysplitbox}
\myparagraph{The Setup phase.}
The setup phase is an offline stage before Alice and Bob run the protocol. In this phase:
\begin{itemize}
    \item Alice and Bob communicate with a trusted third party to obtain cryptographic parameters
    $\ppnew \gets \setupcom(1^\lambda)$ and $\crs \gets \setupzkp(1^\lambda)$.
    \item Bob verifies the authenticity of his dataset $D_B$ with a data authority, which issues a signature attesting to the authenticity of $D_B$.
    \item Alice may optionally perform model distillation locally, then interacts with a model authority that splits $M$ into $C \circ B \circ A$ and certifies that the model split is valid (Sec.~\ref{sec:leakact}).
    \item Alice and Bob jointly fix the non-cryptographic public parameters listed in Table~\ref{tab:public-params}, decided out-of-band before the main protocol.
\end{itemize}

\begin{table}[t]
  \centering
  \footnotesize
  \caption{Public protocol parameters agreed by Alice (model owner) and Bob (data owner) before running PrivaDE.}
  \label{tab:public-params}
  \begin{tabularx}{\columnwidth}{lX}
    \toprule
    Symbol & Description \\
    \midrule
    $m_{\mathsf{CP}}$ 
      & Number of challenges in the CP protocol. \\[0.1em]
    $d$ 
      & Distance threshold used in CP and in the $(d,\delta)$-representativeness condition. \\[0.1em]
    $\delta$ 
      & Allowed outlier ratio in the representativeness guarantee. \\[0.1em]
    $k$ 
      & Size of the representative set $D_R$ (i.e., $|D_R| = k$). \\[0.1em]
    $R$ 
      & Public algorithm for selecting the representative set $D_R$ from $D_B$. \\[0.1em]
    $\ell,\ \mathcal{U},\ \mathcal{D},\ f$ 
      & Public definition of the scoring function: loss $\ell$, 
        uncertainty functional $\mathcal{U}$, diversity functional $\mathcal{D}$,
        and aggregator $f$. \\
    \bottomrule
  \end{tabularx}
\end{table}

This is the only phase that involves a trusted third party. 
The main protocol then proceeds as in Fig.~\ref{fig:proc-privade}. 

\begin{protocolsplitbox}{PrivaDE}{PrivaDE for secure dataset scoring.}{fig:proc-privade}
    \begin{itemize}
        \item[] \textbf{Parties:} Model owner \(\alice\), data owner \(\bob\)
        \item[] \textbf{Private parameters:} Dataset \(D_B\) (held by \(\bob\)); model weights \(\theta_A,\theta_C\) (held by \(\alice\))
        \item[] \textbf{Public parameters:} Commitment parameter \(\ppnew\), representative set size \(k\), distance threshold \(d\), outlier ratio \(\delta\), model architecture \(M\), number of challenges $m_\mathsf{CP}$, model weights $\theta_B$
        \item[] \textbf{Stage 0: Commitments}
            \begin{enumerate}
               \item \(\alice\) computes \((\com_A,r_A)\gets\commit(\ppnew,\theta_A)\) and \((\com_C,r_C)\gets\commit(\ppnew,\theta_C)\), and sends \(\com_A,\com_C\) to \(\bob\).
                \item For each \((x_i,y_i)\in D_B\), \(\bob\) computes \((\com_{x_i},r_{x_i})\gets\commit(\ppnew,x_i)\) and \((\com_{y_i},r_{y_i})\gets\commit(\ppnew,y_i)\), and sends \(\com_{x_i},\com_{y_i}\) to \(\alice\).
            \end{enumerate}

        \item[] \textbf{Stage 1: Representative Set Selection}
            \begin{enumerate}[resume]
                \item \(\bob\) runs the representative set selection algorithm \(\mathcal{R}\) locally to select a subset \(D_R\) of size \(k\) from \(D_B\). It then computes \(d'\) as the \((1-\delta)\)-th percentile of the pairwise distances 
                $\Big\{\min_{b \in D_R^x}\|a-b\| : a \in D_B^x \setminus D_R^x\Big\}$.
                If \(d' > d\), \(\bob\) aborts the protocol. Otherwise, he sends indices $I_R$ to \(\alice\).
                \item \(\alice\) and $\bob$ runs the challenge protocol \(\mathsf{CP}\), where $\alice$ supplies a list of challenge indices $I$ of size m. If the challenge protocol fails, \(\alice\) aborts.
            \end{enumerate}

        \item[] \textbf{Stage 2: Secure Model Inference (Core Component 1)}
            \begin{enumerate}[resume]
               \item \(\alice\) and \(\bob\) run the 2PC protocol \(\Pi_\mathsf{Inference}\) to evaluate block \(A\) on \(D_R^x\). \(\alice\) inputs \((\theta_A,r_A)\); \(\bob\) inputs \(D_R^x\) (and the corresponding opening randomness). \(\bob\) receives the activations \(\arrowvec{a_A}=\{a_{A,i}\}_i\) and per-sample commitments \(\{(\com_{a_i},r_{a_i})\}_i\); both parties obtain \(\{\com_{a_i}\}_i\).
                 \item \(\bob\) locally computes \(\arrowvec{a_B}\gets B_{\theta_B}(\arrowvec{a_A})\). He generates a ZK proof \(\pi_B\) for the language \(L_\textsf{ML-Inf}\) (public weights, hidden inputs) with witness \((\arrowvec{a_A},\{r_{a_i}\}_i)\), and sends \((\arrowvec{a_B},\pi_B)\) to \(\alice\).
                \item \(\alice\) verifies \(\pi_B\); if verification fails, she aborts.
                \item \(\alice\) computes \(\arrowvec{y'}\gets C_{\theta_C}(\arrowvec{a_B})\) and for each \(i\) computes \((\com_{y'_i},r_{y'_i})\gets\commit(\ppnew,y'_i)\). She then produces a ZK proof \(\pi_C\) for \(L_\textsf{ML-Inf}\) (hidden weights, public inputs) with witness \((\theta_C,r_C,y'_1,\ldots,y'_k,r_{y'_1},\ldots,r_{y'_k})\), and sends \(\{\com_{y'_i}\}_{i=1}^k\) and \(\pi_C\) to \(\bob\).
                \item \(\bob\) verifies \(\pi_C\); if verification fails, he aborts.
            \end{enumerate}

        \item[] \textbf{Stage 3: Dataset Scoring (Core Component 2)}
            \begin{enumerate}[resume]
               \item \(\alice\) and \(\bob\) run the maliciously secure 2PC protocol \(\Pi_\mathsf{SubScore}\). \(\alice\) inputs \(\big(y'_1,\ldots,y'_k,\, r_{y'_1},\ldots,r_{y'_k},\, \com_{y'_1},\ldots,\com_{y'_k}\big)\);
                \(\bob\) inputs \(\big(D_B^x,\, y_1,\ldots,y_k,\, r_{y_1},\ldots,r_{y_k},\, \com_{y_1},\ldots,\com_{y_k}\big)\).
                Both parties receive the score \(\phi\).
            \end{enumerate}
    \end{itemize}
\end{protocolsplitbox}

\begin{theorem} \label{thm:privade}
PrivaDE securely realizes \(\Func^k_{\score}\) with malicious security.
\end{theorem}

The proof of Theorem~\ref{thm:privade} will be provided in Appendix~\ref{app:proof-privade}.
\section{Practical Optimizations for PrivaDE}
\label{sec:design}
In this section, we introduce the design techniques that make PrivaDE's secure model inference practical in terms of computation and communication.
\subsection{Model Distillation}
\label{design:md}
Model distillation for modern neural networks, introduced by Hinton et al. \cite{hinton2015distilling}, is widely used to reduce model size and inference cost. The original (teacher) model guides training of a smaller (student) model, typically by matching the teacher's logits.

Distillation is also common in secure inference (e.g., \cite{splithe}) to lower protocol runtime. Empirical studies, such as  \cite{Romero2014FitNetsHF,efficacyKD}, showed that student models with over $10\times$ fewer parameters can retain accuracy comparable to the teacher, depending on task and architecture.

In \textsc{PrivaDE}, we include optional local distillation by the model owner prior to data evaluation, where the choice of student model balances accuracy loss and size reduction. Beyond efficiency, distillation can reduce leakage: limited-capacity students cannot faithfully encode all teacher information~\cite{efficacyKD}. We deliberately select a student size tolerating minor accuracy loss for substantial runtime savings and reduced leakage. Section~\ref{sec:experiments} shows reduced models still support effective data evaluation.

\subsection{Split Model}
\label{design:sm}
\begin{figure}[htbp]
    \centering
    \includegraphics[width=\columnwidth]{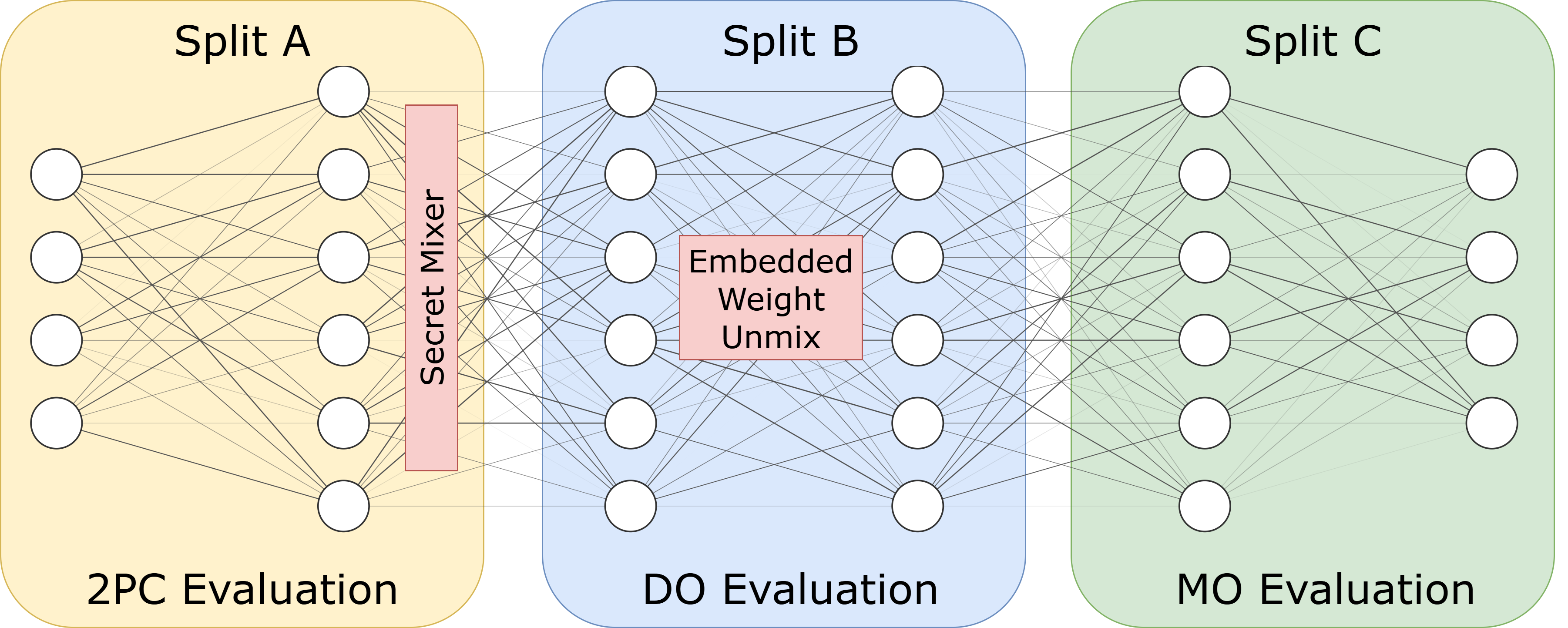}
    \caption{Conceptual diagram of split model in PrivaDE: Split A evaluated in a secure 2PC; split B evaluated by the data owner; split C evaluated by the model owner.}
    \Description{Image showing how a ML model is split into three parts: A, B and C}
    \label{fig:privdesplit}
\end{figure}

Most ML models, particularly neural networks, compose layers: one layer's output feeds the next until the final layer. This enables \emph{split learning}~\cite{splitnn}, partitioning a network at a cut layer. Typically, the data owner trains the front segment and the model owner trains the back segment, coordinating to reduce data exposure.

In secure inference, this yields \emph{split inference} (e.g., \cite{splithe,c2pi}): the front segment executes within a secure protocol, while remaining layers run in the clear, reducing cryptographic workload and latency while limiting input leakage to the (carefully chosen) intermediate activations at the cut layer.

In \textsc{PrivaDE}, we also employ split-model inference, but use a three-way split that factorizes $M$ as $C \circ B \circ A$ (Figure~\ref{fig:privdesplit}):
\begin{enumerate}
\item \textbf{Block $A$.} Contains the initial layers up to (and including) the first activation. We insert a \emph{secret mixer} that permutes $A$'s outputs; this mixer is an invertible linear layer or an invertible $1\times 1$ convolution that mixes channels~\cite{flow}. Block $A$ is evaluated via a maliciously secure 2PC: Alice supplies weights, Bob supplies inputs, and only Bob learns the output of $A$.
\item \textbf{Block $B$.} Comprises the intermediate layers. Its first layer incorporates the inverse mixer to undo $A$'s permutation. The $B$--$C$ cut is chosen by the C2PI boundary-search procedure~\cite{c2pi}. Bob executes $B$ locally after receiving the corresponding weights from Alice.
\item \textbf{Block $C$.} Contains the remaining layers up to the output. Alice executes $C$ after receiving the activations produced by $B$ from Bob.
\end{enumerate}

The $B$--$C$ split follows prior work that balances input-leakage mitigation and runtime. We adopt the C2PI boundary search~\cite{c2pi}, which probes candidate cut layers by attempting input reconstruction from cut-layer activations; for images, a split is accepted when the structural similarity index with the original input falls below $0.3$, which represents the image not being recognizable by human eye.

The $A$--$B$ split further reduces 2PC overhead by restricting cryptographic computation to the minimal front segment (up to the first activation), leaving the remainder in $B$. Because $B$'s weights are sent in the clear to Bob, $A$ prevents him from obtaining a self-sufficient front model that yields reusable embeddings on arbitrary inputs.

Finally, we attach zero-knowledge proofs to the evaluations of $B$ and $C$ to enforce correctness and deter cheating; see Sec.~\ref{sec:protocol} for protocol details.

\subsubsection{Leakage Analysis.}
In our split-model framework, partial leakage of models and activations is unavoidable as it is not protected by zero-knowledge proofs or multiparty computation. Following \cite{c2pi}, we therefore adopt an empirically defined client data privacy model to justify and quantify the leakage.

Before executing PrivaDE, both parties must decide whether the resulting leakages are acceptable according to metrics obtained from experimental attacks; if not, they abort the protocol.

\myparagraph{Leakage of model $B$ to Bob:}
Yosinski et al.\ \cite{featureTrans} show that later layers are more task-specific, while early layers capture generic features. Since neither the weights of $C$ nor the model outputs are revealed, Bob cannot reconstruct the full task model. However, $B\circ A$ may still provide useful embeddings and thus be valuable for fine-tuning on related domains; we therefore need to bound the transferability of $B$.

To repurpose the model, Bob must
\begin{enumerate}
    \item reconstruct $A$ to map raw inputs into $B$'s input space; and
    \item apply transfer learning to $B\circ A$ on new tasks.
\end{enumerate}

We summarize model leakage as a tuple $(\Delta\mathrm{Acc},\textsc{LogME})$, where $\Delta\mathrm{Acc}$ captures the reconstruction quality of $A$ and \textsc{LogME} scores the transferability of $B\circ A$. Alice deems the leakage acceptable or not by comparing $(\Delta\mathrm{Acc},\textsc{LogME})$ to baselines from other marketplace models.

For reconstructing $A$, Bob observes only his inputs and the corresponding outputs of $A$. We assume each contributor issues at most $k$ queries through $A$. A secret mixer (which may vary across runs) applies a random invertible permutation to $A$'s outputs, hindering plug-and-play attacks (e.g., attaching public pretrained fronts to $B$) and collusion by contributors pooling I/O pairs.

Empirical studies \cite{orekondy19knockoff} report that successful extraction typically requires on the order of $10^4$ queries; hence security improves when $k \ll 10^4$. To evaluate robustness at query budget $k$, we run a \emph{reconstruction-under-budget} experiment: sample $\{x_i\}_{i=1}^{k}$ from $D_A$ and train a re-initialized copy $A'$ by minimizing
\[
\min_{\theta}\; \frac{1}{k}\sum_{i=1}^{k} \big\| A'(x_i;\theta) - u_i \big\|_2^2
\;+\; \lambda\,\big\| B\big(A'(x_i;\theta)\big) - B(u_i) \big\|_2^2,
\]
where $\lambda>0$ enforces feature consistency through $B$. After training, we fix $B$ and $C$, replace $A$ by $A'$, and measure top-1 test accuracy. Our metric is
\[
\Delta\mathrm{Acc} \;=\; \mathrm{Acc}\big(C\!\circ\!B\!\circ\!A'\big)\;-\;\mathrm{Acc}\big(C\!\circ\!B\!\circ\!A\big).
\]
A split is deemed robust if $\Delta\mathrm{Acc}$ is substantially negative at practical $k$, indicating that $A$ cannot be effectively reconstructed and $B$ is not easily reused under the stated constraints (cf.\ the privacy criterion in \citealp{c2pi}).

We also assess the transferability of $B\circ A$ (assuming $A$ is known). Following You et al.\ \cite{logme}, we compute \textsc{LogME}, the log marginal evidence that a Bayesian linear regressor explains labels given features. Concretely, we pass labeled images through $B(A(x))$ to obtain features and evaluate \textsc{LogME} on these embeddings; lower scores imply weaker linear transfer and thus lower leakage risk.

\myparagraph{Leakage of activations $B(A(x))$ to Alice:}
\label{sec:leakact}
We quantify input leakage via reconstruction quality, training an inverse model that maps $B(A(x))$ back to $x$. For images, we adopt the distillation-based inverse-network attack (DINA) \cite{c2pi} and report the mean structural similarity index (SSIM) on a test set. SSIM, valued in $[0,1]$, measures perceptual similarity in terms of luminance, contrast, and local structure, with higher scores indicating more recognizable reconstructions. We deem a split privacy-preserving when the mean SSIM $< 0.3$, corresponding to low-fidelity, structurally dissimilar reconstructions.

\paragraph{Justification of the 0.3 SSIM threshold}
Following conventions in prior work on model inversion~\cite{miaattack,c2pi}, reconstructions with SSIM $< 0.3$ are visually unrecognizable. SSIM is calibrated to human perception, and values around 0.3 fall in the ``very poor'' quality regime in subjective studies~\cite{hanafy2018sarjamming,hsieh2018noreference}. We adopt this threshold for direct comparability with C2PI.

\subsection{Representative Set Selection}
\label{design:rss}

Because Bob contributes a dataset $D_B$, scoring every point in $D_B$ is costly and increases leakage: repeated inferences can expose information about Alice’s model and Bob’s data distribution.

To mitigate this, our protocol requires Bob to extract a much smaller \emph{representative subset} $D_R \subset D_B$, and then jointly and securely evaluate Alice’s model only on $D_R$. To ensure correctness even if Bob is malicious, we enforce honest sampling via a cut-and-choose subroutine (Sec.~\ref{subsec:CP}). Similar representative-sampling strategies (without security) are standard in batch active learning~\cite{sener2018activelearningconvolutionalneural} and related ML tasks~\cite{mirzasoleiman2013distributed}. While our protocol allows Bob to choose any algorithm that achieves $(d,\delta)$-representativeness, common methods include $k$-center clustering and submodular maximization~\cite{nemhauser1978analysis,mirzasoleiman2013distributed}.

\subsubsection{Dimension Reduction.}
Selection involves pairwise distances (Def.~\ref{def:rep}), which becomes expensive as $d$ grows. We apply random linear projections~\cite{bingham2001random}:
\[
x' = R\,x,\qquad R\in\mathbb{R}^{m\times d},\; m\ll d,
\]
which preserve distances up to $\varepsilon$-distortion by the Johnson - Lindenstrauss lemma~\cite{johnson1984extensions}. Alice and Bob jointly sample $R$ via $md$ secure coin flips~\cite{blum1981coin}, deriving a shared pseudorandom seed. Bob projects locally and runs representative-selection on $\{x'\}$, reducing computation and communication while retaining fidelity.

\subsubsection{Challenge Protocol for Representative Set Selection}
\label{subsec:CP}

A naive way to securely implement the representative-set algorithm $\mathcal{R}$ is via a maliciously secure two-party computation (2PC). However, because state-of-the-art clustering typically relies on costly Euclidean distance computations, this approach incurs substantial overhead.

As an alternative, we introduce the \emph{challenge protocol} $\mathsf{CP}$, which does not verify the correct execution of $\mathcal{R}$ directly but instead tests whether the resulting subset satisfies $(d,\delta)$-representativeness. A brief overview of $\mathsf{CP}$ was given in Sec.~\ref{subsec:building-blocks}, and the full protocol appears in Fig.~\ref{fig:proc-cp} (App.~\ref{app:cp}). Here, we discuss the theoretical guarantees of $\mathsf{CP}$.

\begin{theorem}
\label{thm:cp}
Let $D_B$ be a dataset of size $n$, and let $D_R\subset D_B$ be a purported $(d,\delta)$-representative subset. For any constant $c>0$, if Alice runs the challenge protocol $\mathsf{CP}$ with $|I|=\lceil c\ln n/\delta\rceil$ uniformly random challenges and all challenges succeed, then with probability at least $1-n^{-c}$, $D_R$ is  $(d,\delta)$-representative.
\end{theorem}

The proof appears in Appendix~\ref{pf:cp}. This theorem shows that $\Theta(\ln n/\delta)$ challenges suffice to verify $(d,\delta)$-representativeness with high confidence. 

\subsection{Audit-Style Cut-and-Choose for ML Inference}
\label{sec:cncML}
While significant research has focused on zero-knowledge proofs (ZKPs) for verifying ML inference (e.g., \cite{zkmlaas,kang2022zkml,zkcnn}), constructing end-to-end proofs remains memory- and computation-intensive.

Taking advantage of the fact that ML models are usually a composition of layers, instead of requiring a full proof, we propose a lightweight protocol that detects incorrect \emph{batched} inference with tunable detection probability (e.g., 80--95\%) by sampling and proving \emph{only a small, random subset} of local layer transitions. The protocol $\mathsf{CnCZK}$ serves as a drop-in replacement for a NIZKPoK of the language $L_\textsf{ML-Inf}$ defined in Sec.~\ref{subsec:privade}.

$\mathsf{CnCZK}$ allows partial verification of the inference process as follows: The prover first commits to the model parameters, inputs, and all intermediate activations of a forward pass using Merkle trees, and the verifier then samples a random subset of inputs ($D_S \subset D_R$) and layers $T \subset \{1,\dots,L\}$ to audit. 
For each sampled position $(i,l)$ (where $i \in D_S$, $l \in T$), the prover provides a zero-knowledge proof that the corresponding commitments (of $\theta_l,a_{l-1,i},a_{l,i}$) and the layer computation $m_l(\theta_l,a_{l-1,i}) = a_{l,i}$ is consistent, where $a_{l,i}$ denotes the activation (output) of the model at layer $l$, on the input $i$.
The full protocol is specified in Fig.~\ref{fig:proc-cnczk} (App.~\ref{app:CnCZK}).

In our experiments (Sec.~\ref{sec:experiments}), we replace full ZKP verifications with the cut-and-choose protocol CnCZK to explore practical privacy–efficiency trade-offs. When all layers and data points are verified, CnCZK is equivalent to the full PrivaDE protocol
in Fig.~\ref{fig:proc-privade}.

\subsubsection{Cheating-detection probability}

Because the scoring protocol aggregates all evaluated points, altering a single point has limited impact on the final score. We therefore consider an adversary that corrupts a fraction $\rho\in(0,1)$ of the $N$ inputs.

\begin{theorem}
\label{thm:cnczkp}
Consider $\mathsf{CnCZK}$ on a model with $L$ layers and $N$ inference points, auditing $m$ points and $s$ layers per audited point (both sampled uniformly without replacement). If an adversary corrupts $N\rho$ points, the probability of detecting cheating is bounded below by
\[
\Pr[\textsf{detect}]
\;\ge\;
\sum_{k=0}^{\min(N\rho,\,m)}
\frac{\binom{N\rho}{k}\binom{N(1-\rho)}{\,m-k\,}}{\binom{N}{m}}
\Bigl[1-\Bigl(\tfrac{L-s}{L}\Bigr)^{k}\Bigr].
\]
\end{theorem}

The proof appears in Appendix~\ref{pf:cnczkp}. As an illustration, with $N=100$, $L=10$, $\rho=0.10$, $m=25$, and $s=6$, the detection probability exceeds $80\%$. This setting audits only $ms/(NL)=150/1000=15\%$ of all layer evaluations, yielding about an $85\%$ reduction relative to checking every layer of every point.

\section{Blockchain Integration for PrivaDE}
\label{sec:blockchain}
In this section, we describe how \textsc{PrivaDE} can be integrated into a blockchain-based decentralized data marketplace.
This is a proof-of-concept; detailed implementation aspects are out of scope.

Using a blockchain brings \emph{four} major advantages. First, commitments and data requests can be posted on-chain, enhancing transparency and guaranteeing input immutability even before evaluation begins. Second, the fairness of MPC within \textsc{PrivaDE}—not typically guaranteed—can be enforced by slashing escrow from dishonest parties. Third, ZKP generation in $\mathsf{CP}$ and $\mathsf{CnCZK}$ can be tied to on-chain payments, deterring overuse and improving efficiency. Finally, data transactions can be integrated with evaluation so that, for example, when the score produced by \textsc{PrivaDE} exceeds a threshold, the transaction is triggered atomically.

\begin{figure}[htbp]
    \centering
    \includegraphics[width=\columnwidth]{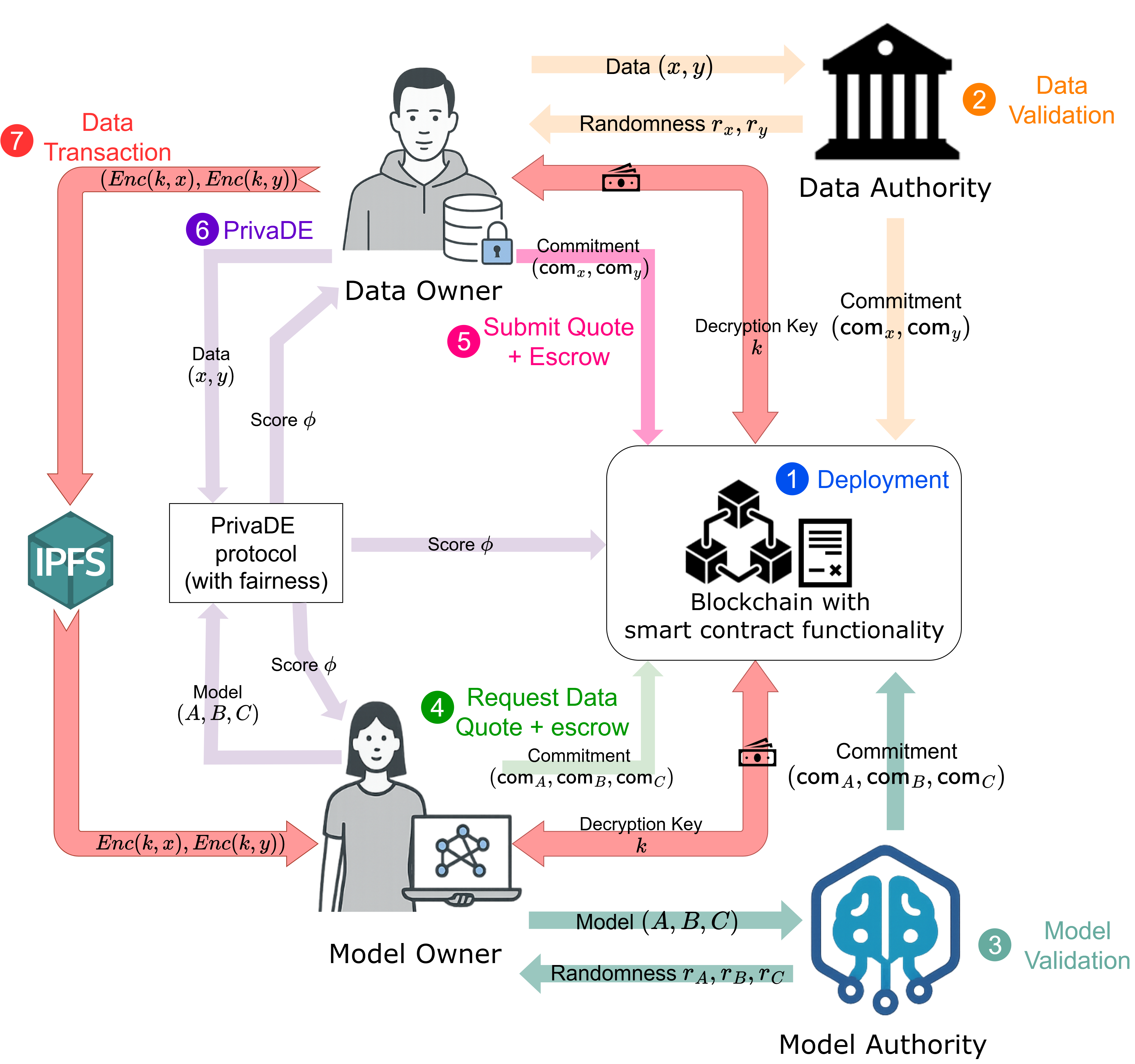}
    \caption{Decentralized Data Marketplace Workflow with PrivaDE integration. Commitments and protocol results are posted on-chain for transparency, and data transaction is seamlessly integrated.}
    \Description{A flowchart illustrating the workflow of a decentralized data marketplace integrated with PrivaDE. The steps include Public Parameter Generation, Data Validation, Model Validation, Request Data Quotes, Submit Data Quotes, Data Evaluation Protocol, and Data Transaction. Each step is represented by a box connected with arrows indicating the flow of the process.}
    \label{fig:ddmarketplace}
\end{figure}

\myparagraph{Step 1: Public Parameter Generation / Smart Contract Deployment}

Before the genesis block, the creator chooses a security parameter $\lambda$ and runs $\ppnew \leftarrow \setupcom(1^\lambda)$. The parameters $(\lambda,\ppnew)$ are embedded as chain parameters.
Once the chain is live, a marketplace smart contract is deployed to record commitments and interactions.

\myparagraph{Step 2: Data Validation}

Before responding to quotes, a data owner must validate their data with a data authority.
The data owner sends $(x,y)$ to the authority for validation.
The authority verifies authenticity and ownership and, if successful, computes
$\com_x,r_x \leftarrow \commit(\ppnew, x)$ and
$\com_y,r_y \leftarrow \commit(\ppnew, y)$,
then submits $(\com_x,\com_y)$ to the marketplace contract from the authority’s contract address (serving as a signature).
The authority notifies the data owner of the outcome and, if successful, returns $(r_x,r_y)$.

\myparagraph{Step 3: Model Validation}

Model owners prepare a model suitable for evaluation (including preprocessing such as distillation and splitting) and submit it to a model authority for validation.
The authority verifies the reconstruction quality at the $B$--$C$ split. If successful, it computes
$\com_{\theta_s}, r_s \leftarrow \commit(\ppnew,\theta_s)$ for $s\in\{A,B,C\}$ and submits \\$(\com_{\theta_A},\com_{\theta_B},\com_{\theta_C})$ on-chain from the \emph{model authority}’s contract address.
The authority notifies the \emph{model owner} and, if successful, returns $(r_A,r_B,r_C)$.

\myparagraph{Step 4: Request Data Quotes}

A model owner requests quotes by calling the marketplace contract, which emits an event that contributors can monitor.
The request includes data requirements, model commitments $(\com_A,$ $\com_B,$ $\com_C)$, and an escrow that is forfeited if the model owner deviates from the protocol.

\myparagraph{Step 5: Submit Data Quotes}

Data contributors respond by submitting their commitments $(\com_x,\com_y)$ and an escrow to the contract.
Submissions must match the authority-published commitments from Step~2; otherwise, the quote is automatically rejected.

\myparagraph{Step 6: Data Evaluation Protocol}

For each accepted quote, the data owner and model owner run \textsc{PrivaDE}, with two blockchain-oriented refinements:
\begin{enumerate}
    \item \textit{MPC fairness.} In $\Pi_\mathsf{SubScore}$, the final protocol message and output can be escrow-enforced: the last-sender must pre-commit (e.g., via an on-chain encrypted blob or hash-locked reference) to the final message/output by a deadline. Failure to comply triggers escrow slashing and compensates the counterparty, preventing an unfair abort after learning $\phi$.
    \item \textit{Payment for ZKP generation.} In $\mathsf{CP}$ and $\mathsf{CnCZK}$, verifiers may over-request proofs to raise the security if there is no cost attached. The contract can require proportional payment from the verifier before proof generation, aligning incentives and reducing prover load.
\end{enumerate}

\myparagraph{Step 7: Data Transaction}

After scoring (potentially across multiple contributors), the model owner proceeds with transactions to selected data owners.
Transactions can use hash-locked exchanges~\cite{poon2016lightning}:
\begin{enumerate}
    \item The data owner uploads an encrypted dataset to IPFS, submits $H(k)$ on-chain, and privately sends the CID and a ZKP attesting that the ciphertext decrypts under $k$;
    \item The model owner retrieves the CID, verifies the ZKP, and escrows payment to the contract;
    \item The data owner reveals $k$ on-chain; the contract checks $H(k)$, releases escrow to the data owner, and the model owner decrypts the data—completing the transaction.
\end{enumerate}

\section{Evaluation}
\label{sec:experiments}
To assess feasibility, we benchmark runtime, memory, communication cost, and scalability by varying model size and dataset complexity. We compare our active data selection against entropy-based sampling and Core-Set~\cite{sener2018activelearningconvolutionalneural}.

Our open-source implementation is publicly available \cite{tpmmthomas_sdv}. We instantiate SMPC using the SPDZ2k protocol within MP-SPDZ~\cite{mpspdz}, employ EZKL~\cite{zkonduit_ezkl} for ZKP-based inference, and implement the challenge protocol in Circom~\cite{circom}.

We validate on MNIST~\cite{lecun2010mnist}, CIFAR-10~\cite{cifar10}, and CIFAR-100, testing three representative models: LeNet (61,706 parameters), ResNet-20 (272,474 parameters), and VGG-8 (4,506,276 parameters).

\subsection{Practicality of PrivaDE}
\label{sec:prac-test}

We implement \textsc{PrivaDE} in a two-party simulation. Each party executes as a separate process on a single host (32-core Intel Core i9, 128~GB RAM), communicating over TCP ports. Loss is cross-entropy $\mathcal{L}_{\mathrm{CE}}$; uncertainty is average prediction entropy; diversity is average feature-wise standard deviation. Representative sets use $k$-center greedy.

An overview of the experimental setup is shown in Table~\ref{tab:privade-overview}. For each dataset we employed models of increasing size, then applied distillation to reduce them. We did not tune model accuracy deliberately, reflecting real-world scenarios where data shortages motivate evaluation of new contributions.

Parameters reflect typical data valuation exchanges. Dataset size $|D_B|$ and model sizes are use-case dependent, while others (representative set size $k$, layers sampled for $\mathsf{CNCZK}$) achieve reasonable security—we require over 85\% cheating detection when adversaries corrupt 15\% of points.

Table~\ref{tab:cp-config-cost} reports $\mathsf{CP}$ protocol results. Setup includes SRS generation, key generation, and compilation; online includes proof generation and verification. Performance is stable across experiments, as $\mathsf{CP}$ runtime is model-size independent; all online times remain below 60 seconds.

The model split (A--B--C) is precomputed and validated (App. \ref{app:model-def}). Table~\ref{tab:model-ab} shows part~A inference within malicious 2PC. Online timings are consistently low (sub-20 seconds), with offline phase dominating due to communication overhead. Offline consists of input-independent randomness generation, hence completable prior to execution.

Inference for parts B and C, executed locally but verified via $\mathsf{CnCZK}$ (Tables~\ref{tab:model-ab},~\ref{tab:model-c-mpc}), show modest resource requirements. Cut-and-choose verification ($\mathsf{CnCZK}$) significantly reduces time, memory, and communication while providing reasonable assurance.

Finally, Table~\ref{tab:model-c-mpc} shows the timings of 2PC data scoring. Runtimes are stable, though CIFAR-100 incurs higher overhead due to its larger output dimension.

Table~\ref{tab:overall-times} summarizes end-to-end runtime. While \textsc{PrivaDE} requires nontrivial resources, it is feasible: smaller models need only about one minute online and less than 15 minutes total. Further optimizations—reducing block A size or lowering $\mathsf{CP}$ and $\mathsf{CnCZK}$ checks—can trade privacy leakage for reduced runtime.

\begin{table*}[t]
  \centering
  \caption{Overview of datasets and model variants tested in section~\ref{sec:prac-test}. Models are tested with increasing size; with distillation that brings 60-90\% reduction in model sizes.}
  \label{tab:privade-overview}
  \scriptsize
  \setlength{\tabcolsep}{5pt}
  \renewcommand{\arraystretch}{1.15}
  \resizebox{\textwidth}{!}{%
  \begin{tabular}{lllcccccc}
    \toprule
    Dataset & \makecell{Original \\Model}& \makecell{Distilled \\Model} & \makecell{Bob Dataset\\Size} & \makecell{Original Model\\Size (\# Params)} & \makecell{Distilled Model\\Size (\# Params)} & Reduction (\%) & \makecell{Teacher\\Accuracy} & \makecell{Student\\Accuracy} \\
    \midrule
    MNIST & LeNet-5 & LeNetXS & 1000 & 61,706 & 3,968 & 93.57 & 0.91 & 0.91 \\
    CIFAR-10 & ResNet20 & LeNet-5 & 1000 & 272,474 & 83,126 & 69.49 & 0.37 & 0.27 \\
    CIFAR-100 & VGG8 & 5-Layer CNN & 1000 & 4,506,276 & 1,727,588 & 61.66 & 0.1 & 0.05 \\
    \bottomrule
  \end{tabular}%
  }
\end{table*}

\begin{table*}[t]
  \centering
  \caption{Overview of parameters and runtime performances of the $\mathsf{CP}$ challenge protocol. Results are stable across different experiments as the protocol is independent of model sizes. }
  \label{tab:cp-config-cost}
  \scriptsize
  \setlength{\tabcolsep}{6pt}
  \renewcommand{\arraystretch}{1.15}
  \resizebox{\textwidth}{!}{%
  \begin{tabular}{lccccccc}
    \toprule
    Dataset & \makecell{Representative \\Set Size } & \makecell{ Data \\ Dimension} & \makecell{ \# Challenges} & \makecell{ $\mathsf{CP}$ Setup \\Time (s)} & \makecell{ $\mathsf{CP}$ Online \\Time (s)} & \makecell{ Memory \\ Usage (MB)} & \makecell{ Communication \\ Overhead (MB)} \\
    \midrule
    MNIST & 50 & 35 & 20 & 456 & 28.0 & 2,579 & 0.0153 \\
    CIFAR-10 & 50 & 50 & 20 & 435 & 50.1 & 2,505 & 0.0154 \\
    CIFAR-100 & 50 & 50 & 20 & 435 & 56.1 & 2,501 & 0.0154 \\
    \bottomrule
  \end{tabular}%
  }
\end{table*}

\begin{table*}[t]
  \centering
  \caption{Model A and Model B inference performance across experiments. 2PC inference of model A incurs significant communication overhead but online runtimes are small. ZKP verification required for model B inference are all within practical runtimes.}
  \label{tab:model-ab}
  \small
  \setlength{\tabcolsep}{6pt}
  \renewcommand{\arraystretch}{1.15}
  \begin{tabular}{lcccccccccc}
    \toprule
    {} & \multicolumn{3}{c}{Model A (2PC)} & \multicolumn{7}{c}{Model B ($\mathsf{CnCZK}$)} \\
    \cmidrule(lr){2-4}\cmidrule(lr){5-11}
    Dataset & \makecell{Online\\(s)} & \makecell{Offline\\(s)} & \makecell{Comm.\\(MB)} & \makecell{\% of Computations\\Verified} & \makecell{Layers\\Sampled} & \makecell{Data points\\Sampled} & \makecell{Setup\\Time (s)} & \makecell{Online\\Time (s)} & \makecell{Memory\\Usage (MB)} & \makecell{Comm.\\(MB)} \\
    \midrule
    MNIST & 4.43 & 193 & 402,343 & 30 & 1/2 & 30/50 & 2.12 & 2.29 & 4,130 & 0.353 \\
    CIFAR-10 & 5.14 & 193 & 346,774 & 25.7 & 3/7 & 30/50 & 5.75 & 6.63 & 5,508 & 1.25 \\
    CIFAR-100 & 20.2 & 761 & 1,355,480 & 30 & 1/2 & 30/50 & 19 & 15 & 10,429 & 6.1 \\
    \bottomrule
  \end{tabular}
\end{table*}

\begin{table*}[t]
  \centering
  \caption{Model C and 2PC Scoring performance across experiments. Model C inference verification scaled with the number of layers sampled, while 2PC scoring scaled with the number of classes.}
  \label{tab:model-c-mpc}
  \small
  \setlength{\tabcolsep}{6pt}
  \renewcommand{\arraystretch}{1.15}
  \begin{tabular}{lcccccccccc}
    \toprule
    {} & \multicolumn{7}{c}{Model C ($\mathsf{CnCZK}$)} & \multicolumn{3}{c}{2PC Data Scoring} \\
    \cmidrule(lr){2-8}\cmidrule(lr){9-11}
    Dataset & \makecell{\% Comp. \\Verified} & \makecell{Layers\\Sampled} & \makecell{Data points\\Sampled} & \makecell{Setup\\Time (s)} & \makecell{Online\\Time(s)} & \makecell{Memory \\Usage (MB)} & \makecell{Comm. \\(MB)} & \makecell{Offline\\Time (s)} & \makecell{Online\\Time (s)} & \makecell{Comm.\\ (MB)} \\
    \midrule
    MNIST & 0.257 & 3/7 & 30/50 & 9.56 & 9.84 & 4,312 & 1.05 & 90.6 & 0.45 & 33,825 \\
    CIFAR-10 & 0.2 & 1/3 & 30/50 & 3.42 & 3.57 & 6,314 & 0.362 & 90.9 & 0.44 & 33,825 \\
    CIFAR-100 & 0.3 & 11/22 & 30/50 & 657 & 779 & 14,962 & 60.9 & 618 & 2.94 & 230,798 \\
    \bottomrule
  \end{tabular}
\end{table*}
\begin{table}[t]
  \centering
  \caption{Overall running times per dataset. Results show a satisfactory online runtime. }
  \label{tab:overall-times}
  \small
  \setlength{\tabcolsep}{6pt}
  \renewcommand{\arraystretch}{1.15}
  \begin{tabular}{lrr}
    \toprule
    Dataset & Online (s) & Total (s) \\
    \midrule
    MNIST & 45.0 & 778 \\
    CIFAR-10 & 65.9 & 794 \\
    CIFAR-100 & 873.0 & 3,360 \\
    \bottomrule
  \end{tabular}
\end{table}

\subsubsection{Scalability Test}
\begin{figure}[htbp]
\centering
\includegraphics[width=\columnwidth]{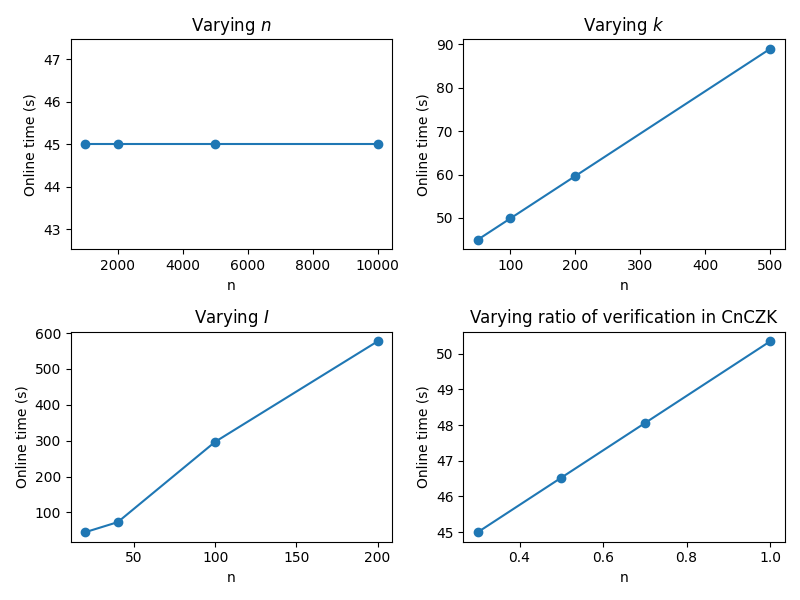}
\caption{Online runtime versus choices of $n$, $k$, $|I|$, and the verification ratio in $\mathsf{CnCZK}$. All parameters except $n$ exhibit approximately linear scaling.}
\Description{Image showing 4 plots arranged in 2 x 2 manner, showing the online runtime in the y-axis, and different parameters in the x-axis. The top-left plot shows the online runtime versus Bob's dataset size n, which is almost flat. The top-right plot shows the online runtime versus representative set size k, which increases linearly. The bottom-left plot shows the online runtime versus number of challenge points |I|, which also increases linearly. The bottom-right plot shows the online runtime versus the percentage of computations verified in CnCZK, which again increases linearly.}
\label{fig:scaling}
\end{figure}

We also evaluate how different parameter choices affect the overall online runtime, and hence the scalability, of PrivaDE. Using the MNIST experiment as a representative case, Figure~\ref{fig:scaling} shows how the online runtime scales with $n$ (Bob's dataset size), $k$ (representative set size), $|I|$ (number of points in the challenge protocol $\mathsf{CP}$), and the ratio of verified computation in $\mathsf{CnCZK}$ for model~B (the behaviour for model~C is analogous).

Bob's dataset size $n$ has almost no effect on the online runtime because only the representative set is used in the scoring protocol; increasing the number of data points does not increase the online phase runtime.

All other parameters scale approximately linearly with the online runtime. The representative set size $k$ directly affects the amount of computation required for 2PC inference of model~A and the scoring component, while $|I|$ and the verification ratio for $\mathsf{CnCZK}$ increase the runtime of their respective parts of the protocol.

Ultimately, the choice of parameters reflects a security--efficiency trade-off. Verifying more computations on more data points increases both parties' confidence in the scoring result and reduces the probability that either party can cheat, but the overall online runtime grows proportionally.

\subsection{Robustness of the scoring algorithm}

The scoring algorithm (Algorithm~\ref{alg:multi-public}) is evaluated against entropy‐based sampling \cite{settles2008analysis} and core‐set sampling \cite{sener2018activelearningconvolutionalneural} on MNIST, CIFAR-10 and CIFAR-100. After normalization and Gaussian blurring, each dataset is split into a pre‐training set (Alice’s initial data), a held‐out test set, and \(N\) contributing sets.

To simulate heterogeneous contributors, the contributing sets are randomly skewed: some contain only one labeled class, some consist of a single data point repeated multiple times, and others contain mixed classes with random label distributions. The scoring algorithm is expected to prioritize datasets that most improve the model and thus boost test accuracy.

Using LeNet for MNIST, ResNet20 for CIFAR-10 and VGG-8 for CIFAR-100 (same setup as above), our framework comprises: representative selection \(\mathcal{S}\) via K‐Center‐Greedy \cite{sener2018activelearningconvolutionalneural}; loss \(\ell=\mathcal{L}_{\mathrm{CE}}\); uncertainty \(\mathcal{U}\) as Shannon entropy; diversity \\\(\mathcal{D}(D_B^x,D_R^x)=\max_{a\in D_B^x\setminus D_R^x}\min_{b\in D_R^x}\|a-b\|\); and scoring \(f(\ell,u,d)=\alpha_1\ell+\alpha_2u+\alpha_3d\) with \(\alpha_1=0.2\), \(\alpha_2=0.1\), \(\alpha_3=0.7\) via grid search.

\begin{figure*}[htbp]
    \centering
    \setlength{\tabcolsep}{1pt}
    \renewcommand{\arraystretch}{0}
    \begin{tabular}{cccc}
        \subcaptionbox{\label{fig:multi_mnist}}{\includegraphics[width=0.3\linewidth]{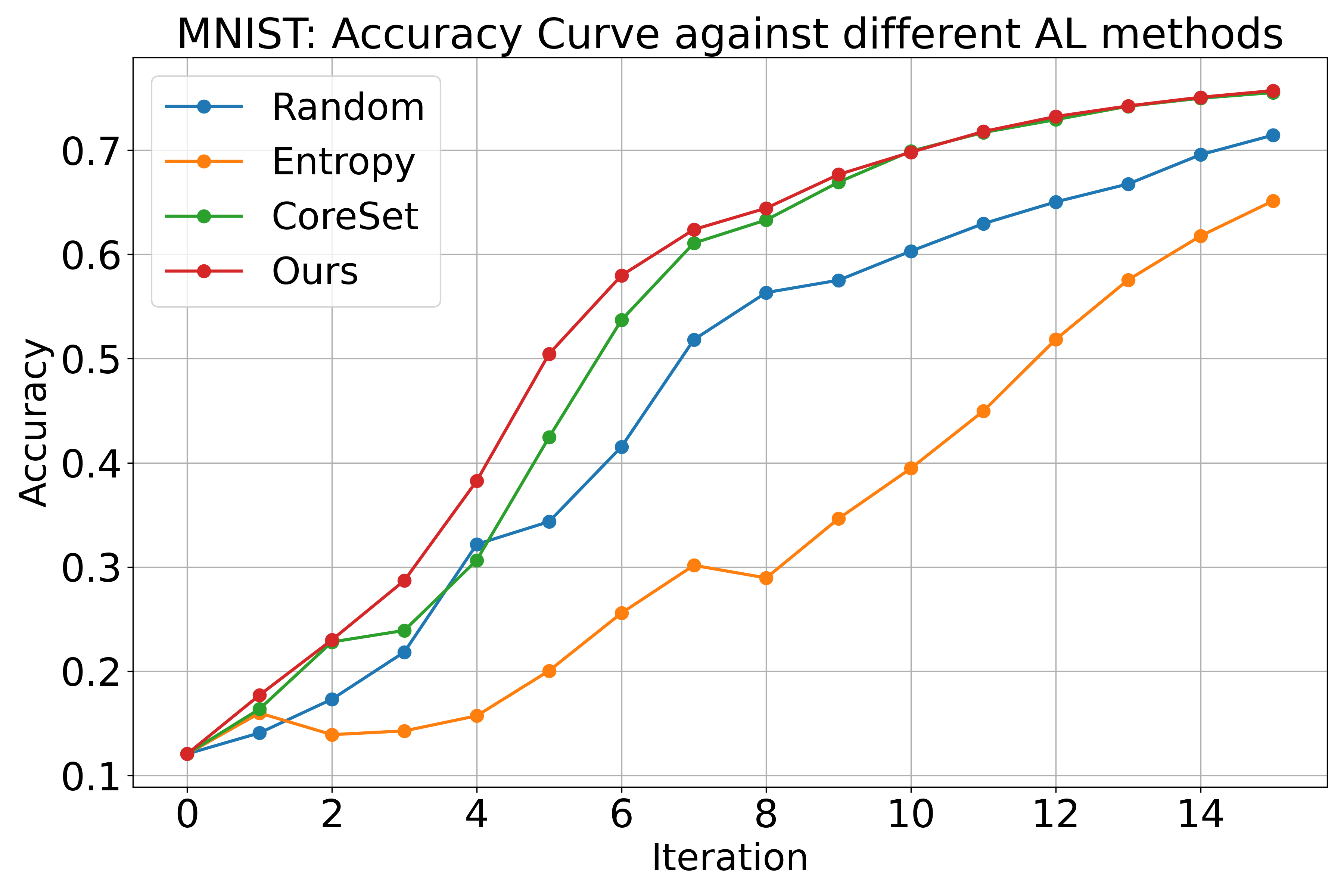}} &
        \subcaptionbox{\label{fig:multi_cifar10}}{\includegraphics[width=0.3\linewidth]{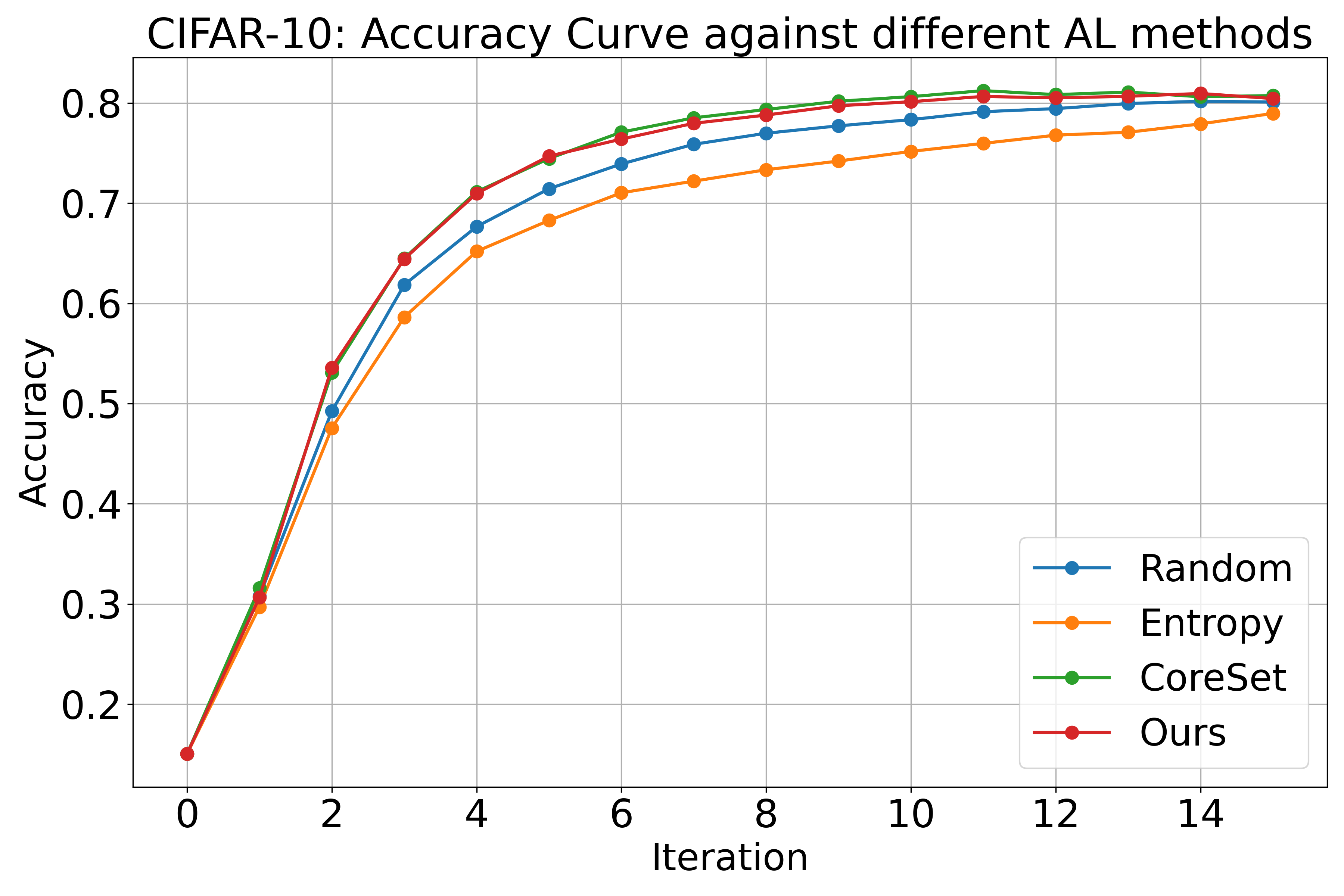}} & 
         \subcaptionbox{\label{fig:multi_cifar100}}{\includegraphics[width=0.3\linewidth]{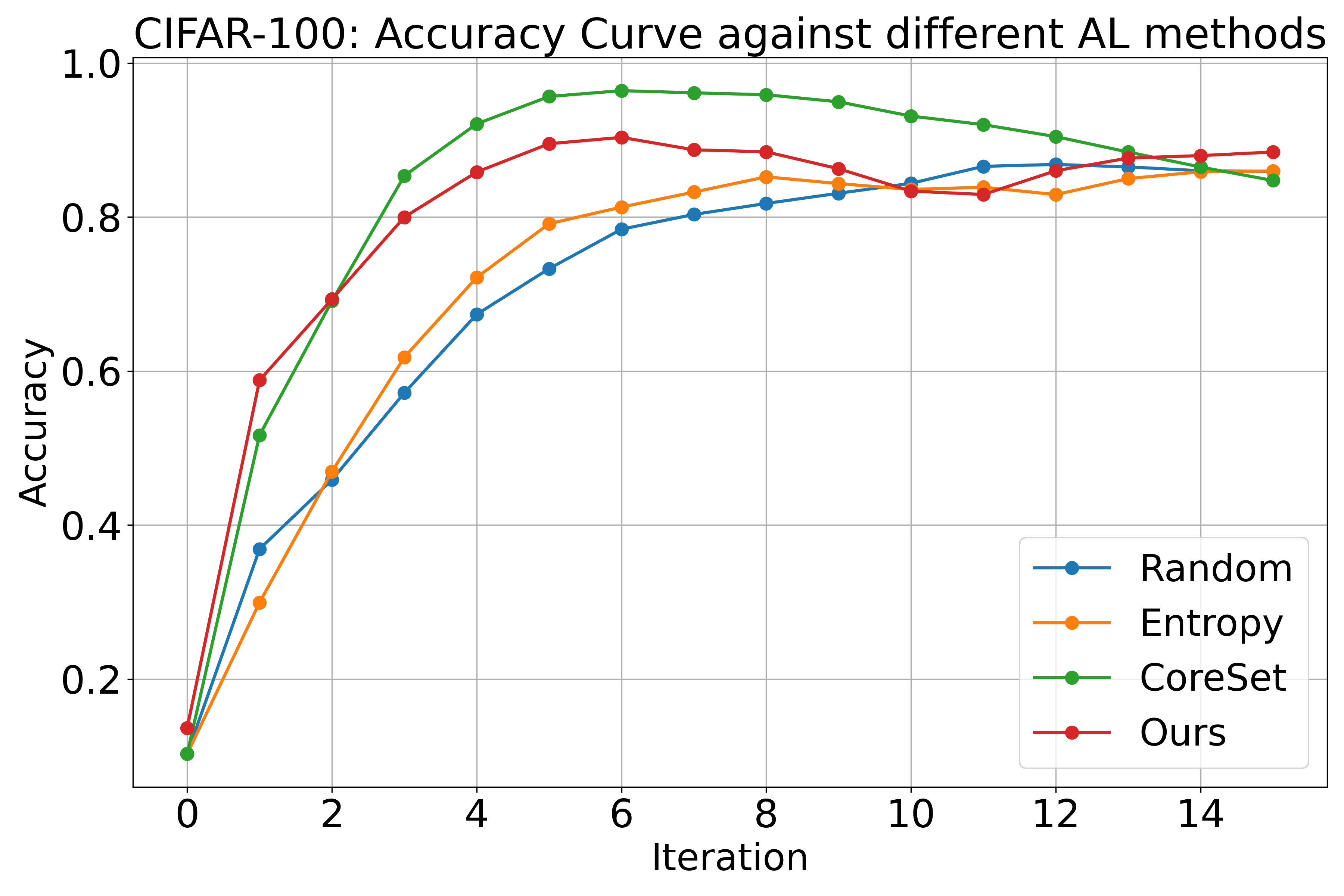}}\\
    \end{tabular}
    \Description{Three plots showing test accuracy versus number of acquired batches for MNIST, CIFAR-10, and CIFAR-100 datasets. In all plots, the multi-point scoring algorithm outperforms random and entropy sampling, and is comparable to core-set sampling.}
    \caption{Robustness of our multi-point scoring algorithm vs. other active learning methods. The results show our method is comparable to core-set and outperforms random and entropy sampling.}
    \label{fig:multi_comp}
\end{figure*}

We simulate sequential acquisition: at each iteration, Alice scores each contributing set, acquires the top batch via our protocol, augments her training data, retrains, and records test accuracy until all \(N\) batches are processed. For baselines, entropy sampling uses mean batch entropy, while core‐set sampling selects ten cluster centers (K‐Center‐Greedy) and scores batches by the maximum sample‐to‐center distance. Accuracy curves are shown in Figure~\ref{fig:multi_comp}.

\myparagraph{Analysis.}
As Figure~\ref{fig:multi_comp} shows, our method consistently outperforms random and entropy sampling. On MNIST (Figure~\ref{fig:multi_mnist}), it yields the largest early gains; on CIFAR-10 (Figure~\ref{fig:multi_cifar10}), it matches core‐set sampling. These results highlight that, under batch constraints, diversity drives early improvements more than uncertainty, and integrating loss, uncertainty, and diversity produces a robust selection criterion across architectures and datasets.

\myparagraph{Effect of using distilled models.}
We further assess the impact of using a distilled model in place of the original teacher model by measuring the Spearman correlation between the scores produced by both. Using the same setup as in the feasibility tests, Table~\ref{tab:corr-mean-sd} reports the results: all teacher–student pairs exhibit high correlation. The only exception is a reduced correlation in the CIFAR-100 setting, which we attribute to the higher class uncertainty and lower baseline accuracy of the original model.

\begin{table}[t]
  \centering
  \caption{Correlation between teacher and student produced scores across 5 runs (mean $\pm$ SD). Results show that all models generally have high correlation, with slightly lower correlation seen in CIFAR-100, possibly attributing to lower accuracy of the initial model.}
  \label{tab:corr-mean-sd}
  \small
  \setlength{\tabcolsep}{6pt}
  \renewcommand{\arraystretch}{1.15}
  \begin{tabular}{lllc}
    \toprule
    Dataset & Teacher & Student & Mean $\pm$ SD \\
    \midrule
    MNIST & LeNet5 & LeNetXS & 0.8742 $\pm$ 0.0592 \\
    CIFAR-10 & resnet20 & LeNet5 & 0.8478 $\pm$ 0.1199 \\
    CIFAR-100 & VGG8 & 5-layer CNN & 0.6285 $\pm$ 0.1158 \\
    \bottomrule
  \end{tabular}
\end{table}
\section{Related Works}
\label{sec:related}

\myparagraph{Privacy-preserving Data Evaluation.}
Prior work has explored privacy-preserving data valuation and selection.

Song et al.~\cite{songDE} propose a data evaluation protocol based on uncertainty sampling, later refined by Qian et al.~\cite{qianDE} with an inner-product-based functional encryption scheme. However, both adopt the semi-honest model, insufficient for blockchain deployments.

Tian et al.~\cite{privatedv} estimate Shapley values via a learned \emph{Data Utility Model} and execute custom MPC. However, this requires sharing data samples and additional labeling to train the utility model; moreover, end-to-end MPC cost remains high (e.g., $\sim$16 hours for 2{,}000 CIFAR-10 images), ill-suited for quick marketplace assessments.

Zhou et al.~\cite{zhoupsmi} and Zheng et al.~\cite{zheng2023csfl} propose secure data valuation schemes for federated learning setups. Zhou et al.~use mutual information to quantify contributions within data coalitions, providing a computationally efficient surrogate for the Shapley value. Zheng et al.~build on Federated Shapley Value (FSV)~\cite{wang2020principledapproachdatavaluation} and design a homomorphic-encryption-based protocol with two central servers to compute FSV. However, both approaches rely on a semi-honest adversarial model and assume trusted servers.

\myparagraph{Data pricing.} Data pricing is usually concerned with setting a price for data in a standard currency. The research in this area adapts economic principles of pricing and markets to data~\cite{zhang2023survey}, and is concerned with the data-specific challenges such as duplication and arbitrage~\cite{chen2019towards,ohrimenko2019collaborative}. These works take the perspective of the broker and consider the seller and buyer's market separately. The actual utility of the data to the consumer is assumed to be determined by simple attributes~\cite{koutris2015query}. 

\myparagraph{Game theoretic fair data valuation.} This topic emerged with the popularity of machine learning and demand for data~\cite{jia2019empirical, ghorbani2020distributional, ghorbani2019data}, and can be seen as an aspect of explainable machine learning~\cite{rozemberczki2022shapleyvaluemachinelearning}. The aim of game theoretic data valuation is to determine fair value for data points (or subsets) in a multiple data contributor setting. These works make use of Shapley Value~\cite{rozemberczki2022shapleyvaluemachinelearning} with its Fairness properties to ensure fair valuation for all. Crucially, in these problems the fairness consideration is between different data contributors, as Shapley Value is a mechanism for determining fair share between multiple players. 
Several associated approaches exist for evaluating data, such as
leave-one-out testing \cite{cook1977detection} and influence function estimation \cite{koh2017understanding, datta2015influence, pruthi2020estimating}, with similar objectives. Shapley value and similar techniques tend to be computationally expensive -- repeatedly training and testing models on different subsets of data -- and require a trusted party to carry out the computation.

\myparagraph{Active learning.} We borrowed several ideas from this area, such as the use of uncertainty and diversity~\cite{settles2009active,dasgupta2011two}. Methods that use both are called hybrid methods. There are algorithms incorporating different scoring values non-linearly~
\cite{ash2020deepbatchactivelearning, yang2017suggestiveannotationdeepactive, zhdanov2019diverseminibatchactivelearning} 
and linearly~\cite{CARDOSO2017313}. In the standard active learning paradigm, loss is not a decision criterion, because loss computation requires the ground truth labels, which are assumed to be expensive, and the algorithms aim to minimize the use of labels. In~\cite{yoo2019learning}, a loss predictor is trained simultaneously with the model to inform the active selection process.

\myparagraph{Zero knowledge proofs in Machine learning}. Researches in this area have been mainly focused on generalizability and efficiency. For generalizability, there are works mapping machine learning operations into suitable zero-knowledge circuit, and translation of non-linear and floating point operations into suitable approximate encryption-friendly circuits (e.g.~\cite{safetynets}). Other researches in this area have approached the efficiency problem by making use of specific features of models and training algorithms~\cite{otti}. However, current zero knowledge proof systems are considered useful mainly for model inference and they are not efficiently supporting the training process. Commonly they affect the model's accuracy or can be too slow for training even moderate sized models.  

Our setting of data scoring using zero knowledge proofs creates some unique challenges and possibilities. It allows the use of labels and loss in ways that are not possible in standard active learning, and ensures the transmission of enough information to make decision without revealing secret data. 

\myparagraph{Secure MPC Inference Methods.} Although zero‐knowledge proofs (ZKPs) protect model weights and guarantee correctness of inference, they do not by themselves ensure privacy of the query input \(x\). An alternative is to employ maliciously‐secure multiparty computation (MPC) under a dishonest–majority model, which provides both confidentiality and integrity in a two‐party setting. The most prominent example here is MD‐ML \cite{mdml}, which optimizes cryptographic primitives to minimize online inference latency. While MD‐ML achieves low absolute latency on small models, its performance degrades sharply with complexity––incurring roughly a 40× slowdown from a Linear SVM to LeNet. 

Many other maliciously‐secure inference frameworks assume an honest–majority among three or more parties, which is incompatible with two‐party scenarios. Representative systems in this category include Helix \cite{helix}, Blaze \cite{blaze}, and SWIFT \cite{swift}. Because these require a strict majority of honest participants, their security and performance guarantees do not directly translate to our setting.
\section{Conclusion}
\label{sec:concl}

Large-scale AI models and training data remain concentrated among major corporations and institutions. Decentralized data marketplaces aim to lower this barrier by establishing trust and incentivizing data sharing, thereby enabling individual data owners and model builders to collaboratively develop advanced models. 

This paper introduced \textsc{PrivaDE}, a practical and efficient protocol for privacy-preserving data evaluation in decentralized marketplaces. \textsc{PrivaDE} operates in the malicious-adversary setting with explicitly bounded leakage, combining representative-set selection, verified split inference, and secure scoring to deliver actionable utility signals. Our experiments demonstrate its feasibility and efficiency.

Future directions include optimizing cryptographic subroutines for lower latency and communication, improving scalability to ever-larger datasets and models, and building fully decentralized (blockchain-based) deployments that integrate data validation, payments, fair MPC, and model training into a cohesive end-to-end workflow.

\begin{acks}
This work is supported by the Input Output Research Hub (IORH) of the University of Edinburgh.
\end{acks}

\bibliographystyle{ACM-Reference-Format}
\bibliography{reference}

\appendix
\section{Subprotocol definitions}
\subsection{The CP Protocol}
\label{app:cp}
The $\mathsf{CP}$ protocol, as mentioned in section~\ref{subsec:CP}, is used to verify the correctness of representative set selection. Its full definition is shown in Fig.~\ref{fig:proc-cp}. 

\begin{protocolsplitbox}{$\mathsf{CP}$}{Challenge Protocol $\mathsf{CP}$ for $D_R$ representativeness verification.}{fig:proc-cp}
    \begin{itemize}
        \item[] \textbf{Parties:} Model owner $\alice$, data owner $\bob$
        \item[] \textbf{Public parameters:} Indices $I_R = \{i_1,\dots,i_k\}$; commitments of $D_B^x$: $\mathsf{COM}_B=\{\com_1,\ldots,\com_n\}$; commitment parameter $\ppnew$; maximum distance $d$; outlier ratio $\delta$
        \item[]
        \begin{enumerate}
            \item $\alice$ samples a set of indices $I\subset[n]$ uniformly at random and sends it to $\bob$.
            \item For each $j\in I$, $\bob$ recovers $x_i\in D_B^x$ and computes $d_i'=\min_{x^r_j\in D_R^x}\|x^r_j-x_i\|$.
            \item For each $i\in I$:
            \begin{itemize}
              \item If $d_i'<d$, $\bob$ sends a ZK proof $\pi_i$ for the language
              \[
                \begin{aligned}
                L_{\mathsf{CP}}
                &= \Bigl\{\,(\ppnew,d,i,\mathsf{COM}_B,I_R)\ \Bigm|\ \exists\, j,x_i,r_i,x_j,r_j\ \text{s.t.}\\
                &\qquad j \in I_R,\quad \|x^r_j - x_i\| \le d,\\
                &\qquad \open(\ppnew,\com_i,x_i,r_i)=1,\quad \open(\ppnew,\com_j,x_j,r_j)=1 \Bigr\}.
                \end{aligned}
                \]
              \item Otherwise, $\bob$ replies $\mathsf{fail}_i$.
            \end{itemize}
            \item $\alice$ verifies each $\pi_i$. If at least $(1-\delta)\,|I|$ proofs succeed, the protocol accepts; otherwise, $\alice$ aborts.
        \end{enumerate}
    \end{itemize}
\end{protocolsplitbox}

\subsection{The CnCZK Protocol}
\label{app:CnCZK}

The $\mathsf{CnCZK}$ protocol, as introduced in Sec.~\ref{sec:cncML}, is used for partial verification of the ML inference process.

We consider two scenarios: \blue{(1) the model weights and outputs are hidden as part of the witness and the data point is public (blue)}; \red{(2) the data point is hidden as part of the witness and the model weights and outputs are public (red)}. For each scenario, read the black text together with the text in the corresponding colour.

\noindent\textbf{Notation.}
Let $M_\theta$ be a fixed $L$-layer network with parameters $\theta$ (with $\theta_\ell$ the weights at layer $\ell$), inputs $\{x_i\}_{i=1}^N$, intermediate activations $a_{\ell,i}$ for $\ell\in\{0,\dots,L\}$ (with $a_{0,i}=x_i$), and outputs $y'_i=a_{L,i}$. Denote the $\ell$-th layer by $m_{\ell}(\theta_\ell;\cdot)$, so $a_{\ell,i}=m_\ell(\theta_\ell; a_{\ell-1,i})$.

\begin{protocolsplitbox}{$\mathsf{CnCZK}$}{Cut-and-choose Protocol $\mathsf{CnCZK}$ for model inference verification.}{fig:proc-cnczk}
    \begin{itemize}
        \item[] \textbf{Parties:} Prover $\alice$, Verifier $\bob$
        \item[] \textbf{Public parameters:} Public parameter $\ppnew$; audit sizes $m$ (checked points) and $s$ (checked layers per point); \blue{the inputs $\{x_i\}_{i=1}^N$}; \red{the model weights $\{\theta_j\}_{j=1}^L$}.
        \item[]
        \begin{enumerate}
  \item[\blue{(1a)}] \blue{\textbf{Model binding.}
        The prover fixes the model parameters $\{\theta_j\}_{j=1}^L$ for the audit window and publishes a \emph{model commitment} as a Merkle root $R_\theta$:
        $R_\theta, \st_m \gets \mtcommit(\ppnew;\theta_1,\ldots,\theta_L)$.}

  \item[\red{(1b)}] \red{\textbf{Data binding.}
        The prover fixes the inputs $\{x_i\}_{i=1}^N$ for the audit window and publishes a \emph{data commitment} as a Merkle root $R_X$:
        $R_X, \st_d \gets \mtcommit(\ppnew;x_1,\ldots,x_N)$.}

  \setcounter{enumi}{1}
  \item \textbf{Compute and commit the full trace.}
        The prover runs a full forward pass on all $N$ inputs, obtaining $\{a_{\ell,i}\}$ and $\{y'_i\}$, and publishes Merkle commitments $\{R_\ell\}_\ell$:
        \[
          R_\ell, \st_\ell \;\gets\; \mtcommit\big(\ppnew;a_{\ell,1},\ldots,a_{\ell,N}\big)
          \quad\text{for all }\ell\in\{0,\ldots,L\},
        \]
        where $a_{0,i}=x_i$ and $a_{L,i}=y'_i$.

  \item \textbf{Challenge sampling.}
        The verifier samples a set $S\subseteq [N]$ of $m$ distinct datapoints (uniform without replacement).
        For each $x_i\in S$, the verifier independently samples a set $T_i\subseteq\{1,\ldots,L\}$ of $s$ distinct layers (uniform without replacement).

  \item \textbf{Prove local correctness on sampled layers.}
        For each $x_i\in S$ and each $\ell\in T_i$, the prover produces a ZKP for the NP language $L_\textsf{CnCZK}$ and sends to the verifier:
        \[
          L_\textsf{CnCZK}
          = \left\{
          \begin{array}{l}
          (\ppnew,i,\ell,\blue{R_\theta}, \\\red{\theta_\ell},R_{\ell-1},R_\ell)
          \end{array}
          : \begin{array}{l}
          \exists\,(\blue{\theta_\ell, \pi_\theta},a_{\ell-1,i},a_{\ell,i},\pi_{\ell-1},\pi_\ell) \ \text{s.t.}\\[2pt]
          m_\ell(\theta_\ell;a_{\ell-1,i}) = a_{\ell,i},\\[2pt]
          \blue{\mtverify(\ppnew,R_\theta,\ell,\theta_\ell,\pi_\ell)=1},\\[2pt]
          \mtverify(\ppnew,R_{\ell-1},i,a_{\ell-1,i},\pi_{\ell-1})=1,\\[2pt]
          \mtverify(\ppnew,R_\ell,i,a_{\ell,i},\pi_\ell)=1
          \end{array}
          \right\}.
        \]
        \blue{Additionally, the prover sends to the verifier a Merkle opening for $R_0$ to show that $x_i$ is included in $R_0$: $\pi \gets \mtopen(\ppnew,\st_0,i)$}

        \red{Additionally, the prover sends to the verifier a ZKP for the input-consistency language $L_{\textsf{in}}$ to ensure the correct inputs are used:
        \[
          L_\textsf{in} = \left\{(\ppnew,R_0,R_X,i):
          \begin{array}{l}
          \exists\,x_i,\pi_1,\pi_2 \ \text{s.t.}\\[2pt]
          \mtverify(\ppnew,R_0,i,x_i,\pi_1)=1,\\[2pt]
          \mtverify(\ppnew,R_X,i,x_i,\pi_2)=1
          \end{array}
          \right\}.
        \]}

  \item \textbf{Verification and decision.}
        The verifier checks all proofs \blue{and merkle openings}. Accept if and only if \emph{all} sampled checks pass; otherwise, reject.
\end{enumerate}
    \end{itemize}
\end{protocolsplitbox}

\section{Proofs of Theorems}

\subsection{Proof of Theorem~\ref{thm:cp}}
\label{pf:cp}

We prove that the challenge protocol $\mathsf{CP}$ detects violations of $(d,\delta)$-representativeness with high probability.

Suppose Bob claims that $D_R$ is $(d,\delta)$-representative of $D_B$, but in reality a fraction $\rho>\delta$ of points in $D_B$ are $(d,D_R)$-outliers (i.e., have distance $>d$ from every point in $D_R$). We will show that with $|I|=\lceil c\ln n/\delta\rceil$ uniform random challenges, Alice detects this cheating with probability at least $1-n^{-c}$.

Let $O\subset D_B$ denote the set of outlier points, so $|O|=\rho n$ with $\rho>\delta$. For each challenge index $i\in I$, either:
\begin{enumerate}
    \item $i$ indexes a non-outlier point, in which case Bob can provide a valid ZK proof $\pi_i$ for language $L_{\mathsf{CP}}$, showing that $i$ has a neighbor in $D_R$ within distance $d$; or
    \item $i$ indexes an outlier point, in which case Bob cannot produce a valid proof (by soundness of the ZK proof system) and must reply $\mathsf{fail}_i$.
\end{enumerate}

Alice accepts if and only if at least $(1-\delta)|I|$ proofs succeed. If the true outlier fraction is $\rho>\delta$, then the expected number of outliers in $I$ is $\rho|I|$. Since each outlier in $I$ causes Bob to fail that challenge, the expected number of failures is $\rho|I|$.

For Alice to erroneously accept, the number of outliers sampled in $I$ must be at most $\delta|I|$ (so that at least $(1-\delta)|I|$ challenges succeed). Let $X$ denote the number of outliers sampled in $I$. Then $X$ follows a hypergeometric distribution with parameters $(n,\rho n,|I|)$, and in the limit of large $n$ (or by a Chernoff-type bound for sampling without replacement), we have
\[
\Pr[X \le \delta|I|] \;\le\; \exp\Bigl(-\frac{(\rho-\delta)^2|I|}{2}\Bigr).
\]

Substituting $|I|=c\ln n/\delta$ and using $\rho>\delta$ (so $\rho-\delta>0$), we obtain
\[
\Pr[X \le \delta|I|] \;\le\; \exp\Bigl(-\frac{(\rho-\delta)^2 c\ln n}{2\delta}\Bigr) \;=\; n^{-\frac{c(\rho-\delta)^2}{2\delta}}.
\]

In particular, when $\rho\ge 2\delta$ (a substantial violation), the exponent satisfies
\[
\frac{c(\rho-\delta)^2}{2\delta} \;\ge\; \frac{c\delta^2}{2\delta} \;=\; \frac{c\delta}{2}.
\]
For any constant $c>2$, this yields a detection probability exponentially close to 1.

More generally, even for $\rho=\delta+\varepsilon$ with small $\varepsilon>0$, we have
\[
\Pr[\text{Alice accepts}\mid \rho>\delta] \;\le\; n^{-\Omega(c)},
\]
so the probability that Alice detects the cheating is
\[
\Pr[\text{detect}] \;=\; 1-\Pr[\text{Alice accepts}\mid \rho>\delta] \;\ge\; 1-n^{-c}.
\]
This completes the proof.

\subsection{Proof of Theorem~\ref{thm:cnczkp}}
\label{pf:cnczkp}
In the worst case, for each corrupted point $i$, the adversary modifies only a single value—either one activation $a_{l,i}$ among the $L\!+\!1$ activations ($l\in\{0,\dots,L\}$) or one layer weight $\theta_j$ among the $L$ weights. Because altering activations can be caught with slightly higher probability (if $l\in\{1,\dots,L-1\}$, detection occurs when either layer $l$ or $l\!+\!1$ is checked), we analyze the stronger adversary that alters only a weight.

Hence, the detection probability for a \emph{single} corrupted point, \emph{conditioned} on that point being audited, is bounded by
\[
p_{\text{single}} \;\ge\; \frac{s}{L}.
\]
If $k$ corrupted points are audited, the probability of detecting at least one of them is
\[
1 - \Pr[\text{no corrupted point detected}] \;=\; 1 - (1-p_{\text{single}})^{k}.
\]
Let $K$ be the number of corrupted points drawn into the audit set $S$; then $K$ follows a hypergeometric distribution with
\[
\Pr[K=k] \;=\; \frac{\binom{N\rho}{k}\binom{N(1-\rho)}{\,m-k\,}}{\binom{N}{m}},\qquad
k=0,1,\dots,\min(N\rho,m).
\]
Therefore,
\[
\begin{aligned}
\Pr[\textsf{detect}]
&= \sum_{k=0}^{\min(N\rho,\,m)} \Pr[K=k]\cdot\bigl[1-(1-p_{\text{single}})^k\bigr] \\
&\ge \sum_{k=0}^{\min(N\rho,\,m)}
\frac{\binom{N\rho}{k}\binom{N(1-\rho)}{\,m-k\,}}{\binom{N}{m}}
\left[1-\left(\tfrac{L-s}{L}\right)^{k}\right],
\end{aligned}
\]
which proves the claim.

\section{Security Definitions}

\subsection{Commitment Schemes}
\label{app:com}

\begin{definition}[Commitment Schemes] \label{def: commitment}
     A \emph{commitment scheme} is a tuple $\Gamma = (\setup, \commit, \open)$ of $\ppt$ algorithms where:
    \begin{itemize} 
        \item $\setup(1^{\lambda}) \rightarrow \ppnew$ takes security parameter $\lambda$ and generates public parameters $\text{pp}$;
        \item $\commit(\ppnew; m) \rightarrow (\com,r)$ takes a secret message $m$, outputs a public commitment $Com$ and a randomness $r$ used in the commitment.
        \item $\open(\ppnew, \com; m, r) \rightarrow b \in \{0,1\}$ verifies the opening of the commitment $C$ to the message $m$ with the randomness $r$.
    \end{itemize}
    
    $\Gamma$ has the following properties:
    \begin{itemize} 
    \item {\bf Binding.} For all $\ppt$ adversaries $\mathcal{A}$, it holds that:
    \[
        \Pr\left[
            b_0 = b_1 \neq 0 \land m_0 \neq m_1 :
            \begin{array}{l}
                \ppnew \leftarrow \setup(1^{\lambda}) \\
                (\com, m_0, m_1, r_0, r_1) \leftarrow \mathcal{A}(\ppnew) \\
                b_0 \leftarrow \open(\ppnew, \com, m_0, r_0) \\
                b_1 \leftarrow \open(\ppnew, \com, m_1, r_1)
            \end{array}
        \right] \leq \negligible(\lambda)
    \]
    \item {\bf Hiding.} For all $\ppt$ adversaries $\mathcal{A}$, it holds that:
    \[
       \left|\Pr\left[
            b_0 = b' :
            \begin{array}{l}
                \text{pp} \leftarrow \setup(1^{\lambda}) \\
                (m_0, m_1, \text{st}) \leftarrow \mathcal{A}(\text{pp}) \\
                b \overset{\$}{\leftarrow} \{0, 1\} \\
                (C_b, r_b) \leftarrow \commit(\text{pp}, m_b) \\
                b' \leftarrow \mathcal{A}(\text{pp}, \text{st}, C_b)
            \end{array}
        \right] - \frac{1}{2} \right| = \negligible(\lambda)
    \]
    \end{itemize}

    \end{definition}

\subsection{Merkle Tree}
\label{app:merkle}
\begin{definition}[Merkle Tree Commitments] \label{def:merkle}
Fix a collision-resistant hash function family $\mathcal{H}=\{H_{\text{pp}}:\{0,1\}^* \to \{0,1\}^\kappa\}$ with public parameter $\text{pp}$ and domain-separation tags $\mathsf{leaf},\mathsf{node}\in\{0,1\}^*$. A (binary) Merkle tree commitment scheme is a tuple
\[
\Delta = (\mtsetup,\ \mtcommit,\ \mtopen,\ \mtverify)
\]
of $\ppt$ algorithms where:

\begin{itemize}
  \item $\mtsetup(1^\lambda) \rightarrow \text{pp}$: samples $\text{pp}$ for $H_{\text{pp}}\leftarrow \mathcal{H}$ 
  
  \item $\mtcommit(\text{pp};\ m_1,\dots,m_n) \rightarrow (\com,\st)$: on inputs $m_1,\dots,m_n$ (the ``leaves''), compute for each $i\in[n]$ a leaf hash
  \[
    \ell_i \;=\; H_{\text{pp}}\bigl(\mathsf{leaf} \,\|\, \mathsf{enc}(i) \,\|\, m_i\bigr)
  \], then iteratively compute internal nodes
  \[
    v \;=\; H_{\text{pp}}\bigl(\mathsf{node} \,\|\, v_L \,\|\, v_R\bigr)
  \]
  up to the root $v_{\mathsf{root}}$. Output the commitment $\com \gets v_{\mathsf{root}}$ and opening state $\st$ containing the tree (or any data sufficient to derive authentication paths).
  
  \item $\mtopen(\text{pp},\st;\ i) \rightarrow \pi_i$: outputs a Merkle \emph{opening} for position $i\in[n]$, namely the authentication path
  \[
    \pi_i \;=\; \bigl( (s_1,b_1),\dots,(s_h,b_h) \bigr),
  \]
  where $s_j$ is the sibling hash at level $j$ and $b_j\in\{L,R\}$ indicates whether the running value is a left or right child; $h=\lceil \log_2 n\rceil$.
  
  \item $\mtverify(\text{pp},\com;\ i,m_i,\pi_i)\rightarrow b\in\{0,1\}$: recompute
  \[
    x_0 \gets H_{\text{pp}}\bigl(\mathsf{leaf} \,\|\, \mathsf{enc}(i) \,\|\, m_i\bigr),\ \ 
    x_{j} \gets 
      \begin{cases}
        H_{\text{pp}}\bigl(\mathsf{node}\,\|\, x_{j-1}\,\|\, s_j\bigr) & \text{if } b_j=L,\\
        H_{\text{pp}}\bigl(\mathsf{node}\,\|\, s_j \,\|\, x_{j-1}\bigr) & \text{if } b_j=R,
      \end{cases}
  \]
  and output $b=1$ iff $x_h=\com$.
\end{itemize}
\end{definition}

Let $\Delta$ be as in Definition~\ref{def:merkle}. Then:

\begin{itemize}
  \item \textbf{Completeness.}
  For all $n$ and $\vec m\!\in\!(\{0,1\}^*)^n$, if 
  $\text{pp}\!\gets\!\mtsetup(1^\lambda)$, 
  $(\com,\st)\!\gets\!\mtcommit(\text{pp},\vec m)$, and 
  $\pi_i\!\gets\!\mtopen(\text{pp},\st;i)$,
  then $\mtverify(\text{pp},\com;i,m_i,\pi_i)=1$ for every $i\in[n]$.

  \item \textbf{Position-Binding.}
  For all $\ppt$ adversaries $\mathcal{A}$, the probability that
  $\mtverify(\text{pp},\com;i,m_0,\pi_0)= \allowbreak \mtverify(\text{pp},\com;i,m_1,\pi_1) \allowbreak =  1$ with $m_0\neq m_1$
  is at most $\negligible(\lambda)$, given 
  $\text{pp}\!\gets\!\mtsetup(1^\lambda)$ and 
  $(\com,i,m_0,\pi_0,m_1,\pi_1)\!\gets\!\mathcal{A}(\text{pp})$.
  (By collision resistance of $H$.)
\end{itemize}

\subsection{Zero-Knowledge Proofs}
\label{app:zk}

\begin{definition}[Non-Interactive Zero-Knowledge proof-of-knowledge Systems] \label{def: NIZK systems}
    A {\em non-interactive zero-knowledge proof-of-knowledge system (NIZKPoK)} for an \NP-language $L$ with the corresponding relation $\relation{L}$ 
    is a non-interactive protocol $\Pi = (\setup, \prover, \verifier)$, where:
    \begin{itemize}
        \item $\setup( 1^{\lambda}) \rightarrow \crs$ takes as the input a security parameter $\lambda$. It outputs a common reference string $\crs$.
        \item $\prover(\crs, x, w) \rightarrow \pi$ takes as the input $\crs$, the statement $x$ and the witness $w$, s.t. $(x, w) \in \relation{L}$. It outputs the proof $\pi$.
        \item $\verifier(\crs, x, \pi) \rightarrow b \in \{0,1\}$ takes as the input $\crs$, $x$ and $\pi$. It outputs 1 to accept and 0 to reject.
    \end{itemize}
    
    $\Pi$ has the following properties:
    \begin{itemize}
    \item {\bf Completeness.} For all $\lambda \in \NN$, and all $(x,w) \in \relation{L}$, it holds that:
    $$\condprob{\verifier(\crs, x, \prover(\crs, x, w)) = 1}{\crs \sample \setup(1^{|x|}, 1^\lambda)} = 1 - \negligible(\lambda)$$
    
    \item {\bf Soundness.} For all $\ppt$ provers $\prover^{\star}$, s.t. for all $\lambda \in \NN$, and all $x\notin L$, it holds that:
    $$\condprob{\verifier(\crs,x,\pi) = 1}{\crs \sample \setup(1^{|x|}, 1^{\lambda}); \pi \sample \prover^{\star}(\crs)} \leq \negligible(\lambda).$$
    
    \item {\bf Zero knowledge.} There exists a $\ppt$ simulator $\simu$ such that 
    for every $(x,w)\in \relation{L}$, the distribution ensembles
    $\{(\crs,\pi): \crs \sample \setup(1^{|x|}, 1^{\lambda});$ $\pi \sample \prover(\crs, x, w)\}_{\lambda \in \NN}$
    and $\{\simu(1^{\lambda},x)\}_{\lambda\in\NN}$
    are computationally indistinguishable.
    
    \item {\bf Proof of Knowledge.} For all $\ppt$ provers $\prover$, there exists an extractor $\mathcal{E}$ such that
    $$\condprob{\mathcal{E}(x,\prover) = w}{(x,w) \in \relation{L}} = 1 - \negligible(\lambda)$$
    \end{itemize}

    \end{definition}

\subsection{Multi-Party Computation}
\label{app:mpc}

Here, we provide a formal definition of secure multiparty computation, with emphasis on the two-party scenario. 

A multiparty protocol is cast by specifying a random process that maps tuples of inputs to tuples of outputs (one for each party). We refer to such a process as a functionality. The security of a protocol is defined with respect to a functionality $f$. In particular, let $n$ denote the number of parties. A (non-reactive) $n$-party functionality $f$ is a (possibly randomized) mapping of $n$ inputs to $n$ outputs. A multiparty protocol with security parameter $\secpar$ for computing the functionality $f$ is a protocol running in time $\poly(\secpar)$ and satisfying the following correctness requirement: if parties $P_1,\dots,P_n$ with inputs $x_1,\dots,x_n$ respectively, all run an honest execution of the protocol, then the joint distribution of the outputs $y_1,\dots,y_n$ of the parties is statistically close to $f(x_1,\dots,x_n)$. 

The security of a protocol (w.r.t.\ a functionality $f$) is defined by comparing the real-world execution of the protocol with an ideal-world evaluation of $f$ by a trusted party. More concretely, it is required that for every adversary $\adversary$, which attacks the real-world execution of the protocol, there exists an adversary $\Sim$, also referred to as the simulator, which can \emph{achieve the same effect} in the ideal-world execution. We denote $\vec{x}=(x_1,\dots,x_n)$.

\myparagraph{Malicious MPC.}
In the real execution for the malicious assumption, the $n$-party protocol $\Pi$ for computing $f$ is executed in the presence of an adversary $\adversary$. The honest parties follow the instructions of $\Pi$. The adversary $\adversary$ takes as input the security parameter $\secpar$, the set $I\subset[n]$ of corrupted parties, the inputs of the corrupted parties, and an auxiliary input $z$. $\adversary$ sends all messages in place of corrupted parties and may follow an arbitrary polynomial-time strategy.

The above interaction of $\adversary$ with a protocol $\Pi$ defines a random variable $\real_{\Pi,\adversary(z),I}(\secpar,\allowbreak\vec{x})$ whose value is determined by the coin tosses of the adversary and the honest players. This random variable contains the output of the adversary (which may be an arbitrary function of its view) as well as the outputs of the uncorrupted parties. 

An ideal execution for a function $f$ proceeds as follows:
\begin{itemize}
	\item \textbf{Send inputs to the trusted party:} As before, the parties send their inputs to the trusted party, and we let $x'_i$ denote the value sent by $P_i$.
	
	\item \textbf{Trusted party sends output to the adversary:} The trusted party computes $f(x'_1,\dots,x'_n)=(y_1,\dots,y_n)$ and sends $\{y_i\}_{i\in I}$ to the adversary.
	
	\item \textbf{Adversary instructs trusted party to abort or continue:} This is formalized by having the adversary send either a continue or abort message to the trusted party. (A semi-honest adversary never aborts.) In the former case, the trusted party sends to each uncorrupted party $P_i$ its output value $y_i$. In the latter case, the trusted party sends the special symbol $\bot$ to each uncorrupted party.
	
	\item \textbf{Outputs:} $\Sim$ outputs an arbitrary function of its view, and the honest parties output the values obtained from the trusted party.
\end{itemize}

In the case of secure two-party computation, where an adversary corrupts only one of the two parties, the above behavior of an ideal-world trusted party is captured by functionality $\function_{\sf 2PC}$, described in Figure~\ref{fig:func-2pc} below.

\begin{functionalitysplitbox}{$\Func_{\sf 2PC}$}{Functionality $\Func_{\sf 2PC}$ for maliciously secure 2PC.}{fig:func-2pc}
\begin{itemize}
    \item[] \textbf{Parameters.} Description of function $f$. 
    Let $\hon \in \{1,2\}$ and $\mal \in \{1,2\}\setminus \{\hon\}$ denote the index of honest party and corrupt party, respectively.
    \item[] 
\begin{enumerate}
    \item $\Func_{\sf 2PC}$ receives input $x_{\hon}$ from honest party $P_{\hon}$ and input $x_{\mal}$ of the corrupt party from simulator $\Sim$.


    \item $\Func_{\sf 2PC}$ computes $y\leftarrow f(x_1,x_2)$ and sends $y$ to $\Sim$.
    
    \item If $\Func_{\sf 2PC}$ receives $\abort$ from $\Sim$, it sends $\abort$ to honest party $P_\hon$. Otherwise, $\Func_{\sf 2PC}$ sends $y$ to honest party $P_{\hon}$.
\end{enumerate}
\end{itemize}
\end{functionalitysplitbox}

As before, the interaction of $\Sim$ with the trusted party defines a random variable $\ideal_{f,\Sim(z),I}(\secpar,\vec{x})$. 
Having defined the real and the ideal world executions, we can now proceed to define the security notion.

\begin{definition}
	\label{def:mpc}
	Let $\secpar$ be the security parameter, $f$ an $n$-party randomized functionality, and $\prot$ an $n$-party protocol for $n\in\NN$. We say that $\prot$ $t$-securely realizes $f$ in the presence of malicious adversaries if for every PPT adversary $\adversary$ there exists a PPT adversary $\Sim$ such that, for any $I\subset[n]$ with $|I|\leq t$, the following quantity is negligible in $\secpar$:
	\[
		|\Pr[\real_{\Pi,\adversary(z),I}(\secpar,\vec{x})=1]-\Pr[\ideal_{f,\Sim(z),I}(\secpar,\vec{x})=1]|,
	\]
	where $\vec{x}=\{x_i\}_{I\in[n]}\in\bits^\ast$ and $z\in\bits^\ast$.
\end{definition}

Note that for two-party computation, $n=2$ and $t=1$ in Definition~\ref{def:mpc}.

\section{Security Proofs}
\subsection{Proof of Thm.~\ref{thm:privade}}
\label{app:proof-privade}

Let $\adversary$ be a PPT adversary. We first consider the case in which $\bob$ (Bob) is corrupted (i.e., $\adversary$ controls $\bob$). The simulator $\Sim_\mathsf{Bob}$ works as follows:
\begin{enumerate}
    \item $\Sim_\mathsf{Bob}$ computes $(\com^A_0, r^A_0) \gets \commit(\ppnew, 0)$ and $(\com^C_0, r^C_0) \gets \commit(\ppnew, 0)$, then invokes $\adversary$ with inputs $\com^A_0, \com^C_0$.
    \item $\adversary$ replies with $\com_{x_i}, \com_{y_i}$ for $i \in [n]$, and the representative set indices $I_R$. 
    \item $\Sim_\mathsf{Bob}$ sends $I_R$ to the trusted party to obtain $I$, and forwards $I$ to $\adversary$.
    \item $\Sim_\mathsf{Bob}$ receives the list of ZKPs $\pi_i$ from $\adversary$. For each $\pi_i$, it verifies the proof and runs the extractor $(j,x_i,r_i,x_j,r_j) \gets \mathcal{E}((\ppnew,d,i,\{\com_{x_i}\},I_R), \adversary)$.
    \item $\Sim_\mathsf{Bob}$ sends $D^x_{R|I} = \{x_j\}_{j=1}^m$ and $D^x_{B|I} = \{x_i\}_{i=1}^m$ to the trusted party. If the trusted party replies with $\abort$, $\Sim_\mathsf{Bob}$ aborts.
    \item $\Sim_\mathsf{Bob}$ invokes $\Sim^\mathsf{Inf}_\mathsf{Bob}$ for subprotocol $\Pi_\mathsf{Inference}$ to obtain $D^x_R, \{r_i\}_i$. It then sends $D^x_R$ to the trusted party, receives activations $\arrowvec{a_A}$, and computes commitments $\com_{a_i}, r_{a_i} \gets \commit(\ppnew,a_i)$. 
    \item It checks each commitment $\open(\ppnew,\com_i,x_i,r_i)$ for $i \in I_R$. If any are invalid, it sends $\abort$ to $\Sim^\mathsf{Inf}_\mathsf{Bob}$. Otherwise, it sends $(\arrowvec{a},\{\com_{a_i},r_{a_i}\}_i)$ to $\Sim^\mathsf{Inf}_\mathsf{Bob}$ and forwards its output to $\adversary$.
    \item $\Sim_\mathsf{Bob}$ receives $(\arrowvec{a_B}, \pi_B)$ from $\adversary$ and $(\theta_B,\arrowvec{a_B}')$ from the trusted party. It verifies $\pi_B$ (for $L_\textsf{ML-Inf}$ with public weights, hidden inputs) using $(\ppnew,\theta_B,\{\com_{a_i}\},\arrowvec{a_B})$ and checks $\arrowvec{a_B} = \arrowvec{a_B}'$. If verification fails, it aborts.
    \item $\Sim_\mathsf{Bob}$ creates $k$ dummy commitments $(\com_{y'_i}, r_{y'_i}) \gets$ \\$ \commit(\ppnew, 0)$ for $i \in [k]$. It then invokes $\Sim_\mathsf{ZK}$ for $L_\textsf{ML-Inf}$ (hidden weights, public inputs) to generate $\pi \gets$\\$\Sim_\mathsf{ZK}(\ppnew,\arrowvec{a_B},\com^C_0,\{\com_{y'_i}\}_i)$. It sends $\pi, \com_{y'_1},\ldots,\com_{y'_k}$ to $\adversary$. If $\adversary$ aborts, it aborts.
    \item $\Sim_\mathsf{Bob}$ invokes $\adversary$ to obtain $(D_B^x,y_1,\ldots,y_k,$ $r_{y_1},\ldots,r_{y_k},$ \\$\com_{y'_1},\ldots,\com_{y'_k})$ for scoring.
    \item $\Sim_\mathsf{Bob}$ sends $(D_B^x,D_R^y)$ to the trusted party and receives score $\phi$.
    \item Finally, $\Sim_\mathsf{Bob}$ checks $\open(\ppnew,\com_{y_i},y_i,r_{y_i})$ for each $\com_{y_i}$. If any fail, it invokes $\Sim^\mathsf{Score}_\mathsf{Bob}$ for $\Pi_\mathsf{SubScore}$ with input $\abort$, otherwise with $\phi$. It outputs whatever $\Sim^\mathsf{Score}_\mathsf{Bob}$ outputs.
\end{enumerate}

We now define the hybrids:
\begin{itemize}
    \item $\hybrid^b_0$: Real-world execution.
    \item $\hybrid^b_1$: Messages for $\Pi_\mathsf{SubScore}$ are computed by $\Sim^\mathsf{Score}_\mathsf{Bob}$. Transition by security of $\Pi_\mathsf{SubScore}$.
    \item $\hybrid^b_2$: The ZKP $\pi$ for model $C$ inference is simulated by $\Sim_\mathsf{ZK}$. Transition by zero-knowledge.
    \item $\hybrid^b_3$: Model weights $\theta_A,\theta_C$ are replaced by 0; inference computed as $y' \gets C_0(x)$ and committed. Transition by hiding of commitments.
    \item $\hybrid^b_4$: Inference of C is skipped; commitments of $y'$ replaced by commitments of 0. Transition by hiding.
    \item $\hybrid^b_5$: Messages for $\Pi_\mathsf{Inference}$ are simulated by $\Sim^\mathsf{Inf}_\mathsf{Bob}$. Transition by security of $\Pi_\mathsf{Inference}$.
\end{itemize}

Each transition is justified by standard properties of the underlying protocols. Hence $\hybrid^b_0 \approx \hybrid^b_5$, and the real and ideal executions are indistinguishable when Bob is corrupted. 

Now consider the case where Alice is corrupted (i.e., $\adversary$ controls $P_1$). The simulator $\Sim_\mathsf{Alice}$ works as follows:
\begin{enumerate}
    \item $\Sim_\mathsf{Alice}$ invokes $\adversary$, which replies with $\com_A, \com_C$. 
    \item For each $i \in [n]$, it creates dummy commitments $(\com_{x_i}, r_{x_i}) \gets \commit(\ppnew, 0)$ and $(\com_{y_i}, r_{y_i}) \gets \commit(\ppnew, 0)$, sending them to $\adversary$.
    \item $\Sim_\mathsf{Alice}$ queries the trusted party for $I_R$ and forwards $I_R$ to $\adversary$.
    \item $\adversary$ replies with indices $I$. $\Sim_\mathsf{Alice}$ forwards $I$ to the trusted party. If it replies with $\abort$, $\Sim_\mathsf{Alice}$ aborts.
    \item For each $i \in I$, it invokes $\Sim_\mathsf{ZK}$ of $\mathsf{CP}$ to generate $\pi_i \gets \Sim_\mathsf{ZK}(\ppnew,d,i,\{com_{x_j}\}_j,I_R)$ and sends all $\pi_i$ to $\adversary$. If $\adversary$ aborts, it aborts.
    \item $\Sim_\mathsf{Alice}$ invokes $\Sim^\mathsf{Inf}_\mathsf{Alice}$ to obtain $(\theta_A,r_A)$ and sends $\theta_A$ to the trusted party. 
    \item It checks $\open(\ppnew,\com_A,\theta_A,r_A)$. If invalid, it sends $\abort$ to $\Sim^\mathsf{Inf}_\mathsf{Alice}$. Otherwise, it creates dummy commitments $\com_{a_i}, r_{a_i} \gets \commit(\ppnew,0)$ and sends $\{\com_{a_i}\}$ to $\Sim^\mathsf{Inf}_\mathsf{Alice}$, forwarding its output to $\adversary$.
    \item $\Sim_\mathsf{Alice}$ sends $\theta_B$ to the trusted party, receives $\arrowvec{a_B}$, and invokes $\Sim'_\mathsf{ZK}$ for $L_\textsf{ML-Inf}$ (hidden inputs, public weights) to generate $\pi \gets \Sim'_\mathsf{ZK}(\ppnew,\theta_B,\{\com_{a_i}\}_i,\arrowvec{a_B})$. It sends $(\arrowvec{a_B},\pi)$ to $\adversary$. If $\adversary$ aborts, it aborts.
    \item $\Sim_\mathsf{Alice}$ queries $\adversary$ for $\pi_C$ and commitments $\com_{y'_1},\ldots,\com_{y'_k}$. It verifies $\pi_C$; if invalid, aborts. 
    \item It extracts $\theta_C$ and $\arrowvec{y'}$ openings from the ZKP, sends $\theta_C$ to the trusted party, and receives score $\phi$.
    \item Finally, it checks $\open(\ppnew,\com_{y'_i},y'_i,r_{y'_i})$ for each $i$. If any fail, it invokes $\Sim^\mathsf{Score}_\mathsf{Alice}$ for $\Pi_\mathsf{SubScore}$ with $\abort$, otherwise with $\phi$. It outputs whatever $\Sim^\mathsf{Score}_\mathsf{Alice}$ outputs.
\end{enumerate}

The hybrids are:
\begin{itemize}
    \item $\hybrid^a_0$: Real-world execution.
    \item $\hybrid^a_1$: Messages for $\Pi_\mathsf{SubScore}$ are simulated by $\Sim^\mathsf{Score}_\mathsf{Alice}$. Transition by security of $\Pi_\mathsf{SubScore}$.
    \item $\hybrid^a_2$: ZKP for model B inference replaced by $\Sim'_\mathsf{ZK}$. Transition by zero-knowledge.
    \item $\hybrid^a_3$: Activations from model $a$ replaced by 0. Transition by hiding.
    \item $\hybrid^a_4$: Messages for $\Pi_\mathsf{Inference}$ simulated by $\Sim^\mathsf{Inf}_\mathsf{Alice}$. Transition by security of $\Pi_\mathsf{Inference}$.
    \item $\hybrid^a_5$: ZKP for challenge protocol $\mathsf{CP}$ simulated by $\Sim_\mathsf{ZK}$. Transition by zero-knowledge.
    \item $\hybrid^a_6$: Identical to ideal world except data points $(x,y)$ replaced by 0. Transition by hiding.
\end{itemize}

Transitions are justified as above. Thus the adversary’s view is indistinguishable from the ideal execution. This completes the proof that PrivaDE securely computes \(\Func_{\score}\).

\section{Model Definitions}
\label{app:model-def}

Table~\ref{tab:lenetxs} shows the LeNetXS model (used with MNIST dataset), table~\ref{tab:lenet5} shows the LeNet5 model (used with CIFAR-10) and table~\ref{tab:cnn5} shows the 5-layer CNN model (used with CIFAR-100).

\begin{table}[H]
\centering
\caption{LeNetXS architecture with split bands.}
\label{tab:lenetxs}
\setlength{\tabcolsep}{5pt}
\renewcommand{\arraystretch}{1.15}
\small
\begin{tabularx}{\linewidth}{@{}p{3mm} r l l r@{}}
\toprule
& \# & Layer (type) & Output Shape & Param \# \\
\midrule
\bandA \textbf{A} & 1  & Conv2d-1             & [-1, 3, 24, 24] & 78 \\
\bandA            & 2  & ReLU-2               & [-1, 3, 24, 24] & 0 \\
\bandB \textbf{B} & 3  & AvgPool2d-3          & [-1, 3, 12, 12] & 0 \\
\bandB            & 4  & Conv2d-4             & [-1, 6, 8, 8]   & 456 \\
\bandC \textbf{C} & 5  & ReLU-5               & [-1, 6, 8, 8]   & 0 \\
\bandC            & 6  & AvgPool2d-6          & [-1, 6, 4, 4]   & 0 \\
\bandC            & 7  & AdaptiveAvgPool2d-7  & [-1, 6, 4, 4]   & 0 \\
\bandC            & 8  & Flatten-8            & [-1, 96]        & 0 \\
\bandC            & 9  & Linear-9             & [-1, 32]        & 3,104 \\
\bandC            & 10 & ReLU-10              & [-1, 32]        & 0 \\
\bandC            & 11 & Linear-11            & [-1, 10]        & 330 \\
\bottomrule
\end{tabularx}
\end{table}

\begin{table}[H]
\centering
\caption{LeNet5 architecture with split bands.}
\label{tab:lenet5}
\setlength{\tabcolsep}{5pt}
\renewcommand{\arraystretch}{1.15}
\small
\begin{tabularx}{\linewidth}{@{}p{3mm} r l l r@{}}
\toprule
& \# & Layer (type) & Output Shape & Param \# \\
\midrule
\bandA \textbf{A} & 1  & Conv2d-1        & [-1, 6, 28, 28]  & 156 \\
\bandA            & 2  & ReLU-2          & [-1, 6, 28, 28]  & 0 \\
\bandB \textbf{B} & 3  & AvgPool2d-3     & [-1, 6, 14, 14]  & 0 \\
\bandB            & 4  & Conv2d-4        & [-1, 16, 10, 10] & 2,416 \\
\bandB            & 5  & ReLU-5          & [-1, 16, 10, 10] & 0 \\
\bandB            & 6  & AvgPool2d-6     & [-1, 16, 5, 5]   & 0 \\
\bandB            & 7  & Flatten-7       & [-1, 400]        & 0 \\
\bandB            & 8  & Linear-8        & [-1, 120]        & 48,120 \\
\bandB            & 9  & ReLU-9          & [-1, 120]        & 0 \\
\bandC \textbf{C} & 10 & Linear-10       & [-1, 84]         & 10,164 \\
\bandC            & 11 & ReLU-11         & [-1, 84]         & 0 \\
\bandC            & 12 & Linear-12       & [-1, 10]         & 850 \\
\bottomrule
\end{tabularx}
\end{table}

\begin{table}[H]
\centering
\caption{5-layer CNN architecture with split bands.}
\label{tab:cnn5}
\setlength{\tabcolsep}{5pt}
\renewcommand{\arraystretch}{1.10}
\scriptsize 
\begin{tabular*}{\linewidth}{@{\extracolsep{\fill}} p{3mm} r l l r @{}}
\toprule
& \# & Layer (type) & Output Shape & Param \# \\
\midrule
\bandA \textbf{A} & 1  & Conv2d-1        & [-1, 32, 32, 32]  & 896 \\
\bandA            & 2  & ReLU-2          & [-1, 32, 32, 32]  & 0 \\
\bandB \textbf{B} & 3  & BatchNorm2d-3   & [-1, 32, 32, 32]  & 64 \\
\bandB            & 4  & MaxPool2d-4     & [-1, 32, 16, 16]  & 0 \\
\bandC \textbf{C} & 5  & Conv2d-5        & [-1, 64, 16, 16]  & 18,496 \\
\bandC            & 6  & ReLU-6          & [-1, 64, 16, 16]  & 0 \\
\bandC            & 7  & BatchNorm2d-7   & [-1, 64, 16, 16]  & 128 \\
\bandC            & 8  & MaxPool2d-8     & [-1, 64, 8, 8]    & 0 \\
\bandC            & 9  & Conv2d-9        & [-1, 128, 8, 8]   & 73,856 \\
\bandC            & 10 & ReLU-10         & [-1, 128, 8, 8]   & 0 \\
\bandC            & 11 & BatchNorm2d-11  & [-1, 128, 8, 8]   & 256 \\
\bandC            & 12 & MaxPool2d-12    & [-1, 128, 4, 4]   & 0 \\
\bandC            & 13 & Conv2d-13       & [-1, 256, 4, 4]   & 295,168 \\
\bandC            & 14 & ReLU-14         & [-1, 256, 4, 4]   & 0 \\
\bandC            & 15 & BatchNorm2d-15  & [-1, 256, 4, 4]   & 512 \\
\bandC            & 16 & MaxPool2d-16    & [-1, 256, 2, 2]   & 0 \\
\bandC            & 17 & Conv2d-17       & [-1, 512, 2, 2]   & 1,180,160 \\
\bandC            & 18 & ReLU-18         & [-1, 512, 2, 2]   & 0 \\
\bandC            & 19 & BatchNorm2d-19  & [-1, 512, 2, 2]   & 1,024 \\
\bandC            & 20 & MaxPool2d-20    & [-1, 512, 1, 1]   & 0 \\
\bandC            & 21 & Flatten-21      & [-1, 512]         & 0 \\
\bandC            & 22 & Dropout-22      & [-1, 512]         & 0 \\
\bandC            & 23 & Linear-23       & [-1, 256]         & 131,328 \\
\bandC            & 24 & ReLU-24         & [-1, 256]         & 0 \\
\bandC            & 25 & Dropout-25      & [-1, 256]         & 0 \\
 \bandC            & 26 & Linear-26       & [-1, 100]         & 25,700 \\
\bottomrule
\end{tabular*}
\end{table}

\end{document}